\documentclass[11pt]{article}
\pdfoutput=1
\usepackage[top=2cm, bottom=2cm, left=2cm, right=2cm]{geometry}
\usepackage{jcappub}
\usepackage{float}
\usepackage{graphicx}
\usepackage{rotating}
\usepackage{color}
\usepackage{amsmath,bm}
\usepackage{hyperref}

\definecolor{purple}{RGB}{160,0,160}
\definecolor{plotpink}{RGB}{205,0,180}
\definecolor{plotcyan}{RGB}{0,215,215}
\definecolor{plotblue}{RGB}{0,0,235}
\definecolor{plotorange}{RGB}{245,140,0}
\definecolor{plotgreen}{RGB}{30,130,0}
\definecolor{plotred}{RGB}{240,0,0}
\definecolor{darkgreen}{RGB}{0,170,0}

\title{Handling the uncertainties in the Galactic Dark Matter distribution for particle Dark Matter searches}
\author[a]{Maria Benito,}
\author[b,c]{Alessandro Cuoco,}
\author[a]{and Fabio Iocco}
\affiliation[a]{ICTP--South American Institute for Fundamental Research, and Instituto de F\'isica Te\'orica / Universidade Estadual Paulista (UNESP), Rua Dr. Bento Teobaldo Ferraz 271, 01140-070 S\~{a}o Paulo, SP Brazil}
\affiliation[b]{Institute for Theoretical Particle Physics and Cosmology,
RWTH Aachen University, 52056 Aachen, Germany}
\affiliation[c]{Univ. Grenoble Alpes, USMB, CNRS, LAPTh, F-74940 Annecy, France}

\emailAdd{mariabenitocst@gmail.com}
\emailAdd{cuoco@physik.rwth-aachen.de}
\emailAdd{fabio.iocco.astro@gmail.com}

\abstract{
In this work we characterize the distribution of Dark Matter (DM) in the Milky Way (MW),
and its uncertainties, adopting the well known ``Rotation Curve'' method.
We perform a full marginalization  over 
the uncertainties of the Galactic Parameters and over the lack of knowledge on the
morphology of the baryonic components of the Galaxy.
The local DM density $\rho_0$ is
constrained to the range $0.3 - 0.8$ GeV/cm$^3$ at the $2 \sigma$ level, and
has a strong positive correlation to $R_0$,  the local distance from the Galactic Center.
The not well-known value of $R_0$  is thus, at the moment, a major limitation in
determining $\rho_0$.
Similarly, we find that the inner slope of the DM profile,  $\gamma$, is very weakly constrained, 
showing no preference
for a cored profile ($\gamma \simeq 0$) or a cuspy one ($\gamma \simeq[1.0,1.4]$).
Some combination of parameters can be, however, strongly constrained. For example
the often used standard $\rho_0=0.3$ GeV/cm$^3$, $R_0=8.5$ kpc is
excluded at more than 4 $\sigma$. 
We release the full likelihood of our analysis in a tabular form 
over a multidimensional grid in the parameters characterizing the DM distribution,
namely the scale radius $R_s$, the scale density $\rho_s$, the inner slope of the profile $\gamma$,
and $R_0$.
The likelihood can be used to include the effect of the DM distribution uncertainty 
on the results of searches for an indirect DM signal  in gamma-rays or neutrinos,
from the Galactic Center (GC), or the Halo region surrounding it.
As one example, we study the case of the GC excess in gamma rays.
Further applications of our tabulated uncertainties in the DM distribution involve
local DM searches, like direct detection and anti-matter observations, or global fits combining local and
GC searches.}


\arxivnumber{}

\begin{document}

\begin{flushright}
TTK-18-37\\
LAPTH-051/18\\
\end{flushright}
\maketitle

\section{Introduction}\label{sec:intro}

The distribution of Dark Matter (DM) within the Galactic Halo
is one crucial ingredient in direct and indirect particle DM searches.
Typically, the DM Halo is assumed to be spherically symmetric and
various analytical parameterizations for the radial profile are
adopted, e.g. Navarro-Frenk-White (NFW~\cite{Navarro:1996gj}), Burkert~\cite{Burkert:1995yz}, Einasto~\cite{Einasto1965} or generalized NFW. 
The uncertainty arising from our lack of knowledge over the DM Halo shape is thus often taken into account by 
the bracketing of benchmark models: to account for the uncertainty in the DM
distribution in the inner Galaxy a cored profile, like Burkert, and a cuspy one,
like NFW, are often adopted as benchmark, and the results shown for the extreme cases. 
Still, further uncertainties are present, like the overall normalization of the profile,
often expressed as the DM density $\rho(R)$ at $R_0$ --the Solar system distance
from the Galactic Center-- and referred to as the local DM density $\rho_0$,
or the overall scale radius of the profile $R_s$.
The quantity $R_0$ itself is also affected by observational uncertainties, which propagate into
the global profile uncertainty.

Taking into account in a self-consistent way all the above uncertainties is not
straightforward, also because strong correlations among the various parameters
are present.
Several analyses constraining the DM Halo distribution and the related parameters have been performed to date, 
e.g.~\cite{Caldwell:1981,Iocco2011, Iocco2015, Pato:2015dua, Sofue:2008wt, Sofue2012, CatenaUllio2010, Nesti:2013uwa, Huang2016, McMillan2011, McMillan2017, DeBoer2009, Xue2008, Bovy2009, BovyAPOGEE2012},
but the correlation matrix is not typically provided, thus making the full results not promptly usable from the reader.

The main goal of this work is to fill this gap and to provide an account as comprehensive as possible of
the uncertainties of both observational and modelization nature. 
We use the well-known method
of Galactic ``Rotation Curve'' to derive the constraints on the parameters
of the Galactic DM Halo  taking the above--mentioned uncertainties and dependencies into account
and we provide the full likelihood, 
so that the interested readers can use these constraints and apply them to their
own analysis. 
Possible applications include analyses of a DM signal, or upper limits,
from the Galactic Center, or its surrounding. Typical cases are searches for
a gamma-ray signal, in the form of lines~\cite{Abdalla:2016olq,Abdallah:2018qtu,Ackermann:2013uma,Ackermann:2015lka},
or smooth spectra either at GeVs (see e.g.~\cite{Calore:2014xka,TheFermi-LAT:2017vmf}), or at TeV
 with Cherenkov telescopes~\cite{Abdallah:2016ygi,HESS:2015cda,Acharya:2017ttl, Doro:2012xx}.
Analogous searches for DM signal from the GC can be performed 
in neutrinos~\cite{Aartsen:2018mxl, Aartsen:2017ulx, Aartsen:2015xej, Adrian-Martinez:2015wey}.
In all these cases the amount of DM
signal in the GC region is proportional to the so-called $\mathcal{J}$-factor, which is a function of the 
properties of the DM Halo.  The uncertainty in the $\mathcal{J}$-factor can then be taken into account
through the likelihood provided in this work\footnote{The likelihood profile from the full data-driven analysis presented in this work is available at \href{https://github.com/mariabenitocst/UncertaintiesDMinTheMW}{https://github.com/mariabenitocst/UncertaintiesDMinTheMW}.}.
Sometimes extra-constraints are also needed, for example one would
like to restrict the slope of the DM Halo near the GC to a certain range of values, or apply
a specific prior on the local DM density. The advantage of providing the full
likelihood is that extra-priors can be easily included just introducing them
as extra multiplication factors in the likelihood, so that to use a new likelihood which
includes all the desired information. 
Similar considerations can be applied to local searches, for example DM searches
with antiprotons, e.g., \cite{Cuoco:2016eej,Giesen:2015ufa,Fornengo:2013xda,Donato:2003xg,Hooper:2014ysa,Bringmann:2006im} or antimatter in general (antinuclei, positrons)~\cite{Gaggero:2018zbd}, whose result is  mostly sensitive to the local DM density
but also to the shape of the DM profile within few kpc of the Solar System.

Finally, it is worth noticing here that direct searches are sensitive to
the local DM density, and to the local DM velocity distribution~\cite{Bhattacharjee2013, Bozorgnia2017, Bozorgnia2016, Fairbairn2013, Fornasa2014, Frandsen2012, Green2017, Green2012, Kuhlen2010, Peter2010, Pillepich2014, Vogelsberger2009}.
Our results can thus be also used to investigate the astrophysical uncertainties
in direct DM searches, although in this study we will only touch upon the subject.


\section{Setup and Data}
\label{sec:setup}

\par To derive constraints on the distribution of DM in the Galaxy we use the method commonly referred to as the
``Rotation Curve''.
This method, largely adopted in the literature to determine the DM existence and distribution 
in rotation supported disc galaxies and in particular in  our own 
e.g.~\cite{Caldwell:1981,Iocco2011, Iocco2015, Pato:2015dua, Sofue:2008wt, Sofue2012, CatenaUllio2010, Nesti:2013uwa, Huang2016, McMillan2011, McMillan2017, DeBoer2009, Xue2008, Bovy2009, BovyAPOGEE2012},
relies on the use of two main elements:
on one hand the actual, observed gravitational potential, which is inferred through the motion of appropriate
tracers of the circular velocity of the disk (the Rotation Curve); on the other hand the potential expected from the baryonic component (gas and stars).
The mismatch between the two (if any) is fit with the expectation (in Newtonian approximation) from a spherical 
DM profile, typically some form of broken power-law.
This is opposed to the ``local'' method, which, instead relies on local observables only,
typically the position and motions of stars within few kpc from the Solar system to infer
the local gravitational potential and DM density, 
e.g.~\cite{Bienayme2014, Bovy2012, Garbari2012, McKee2015, Salucci2010, Sivertsson2018, XiaLAMOST2016, Zhang2013}.
For a review and discussion of both methods see \cite{Read2014}.

\subsection{The observed rotation curve}
\label{sec:data}
As tracer of the total gravitational potential, or Rotation Curve (RC), 
we adopt the data from the {\tt galkin} compilation~\cite{Iocco2015, Pato:2017yai}.
The compilation contains data up to  Galactocentric
radii of $R \sim$ 20 kpc and includes the kinematics of gas, stars and masers for a total of 2780
measurements collected from the literature. More details are reported in the original publications~\cite{Iocco2015, Pato:2017yai},
with extensive studies about the possible source of systematics described in the Supplementary Material of \cite{Iocco2015}.
 {\tt galkin} uses as input  the local standard of rest $(U, V, W)_{\odot}$ (see below), $R_0$, the Solar system distance
 from the Galactic Center, and   $V_0$, the local circular velocity
 to self-consistently transform determinations from different observations into a point on the RC \mbox{$(R_i\pm \delta R_i ,  \omega_i \pm \delta\omega_i)$} with respective errors.  
 The data-points on the RC can be expressed in term of circular velocity $V$, or, circular angular velocity $\omega$,
 which is typically more convenient since in the latter case $R$ and $\omega$ errors are uncorrelated,
 contrary to the former. 
Since $R_0$ is a parameter which we vary in our analysis, the RC curve is self-consistently
updated when $R_0$ is changed. 

To trace the RC from 20 kpc up to $\sim$100 kpc, stellar dispersion data are typically used.
Nonetheless,  since these stellar tracers are not in circular orbit,
the use of these data requires further assumptions, for example about the virialization
of the system and the velocity anisotropy. We prefer to conservatively limit ourselves to the use
of tracers in circular motion such as those contained in the recent compilation
{\tt galkin} --though they imply an intrinsic limitation to the innermost Galaxy
-- and postpone an accurate study with stellar tracers to future work.

\subsection{Baryonic morphology}
\label{sec:bmorphology}
The visible (baryonic) component of the Milky Way, is typically separated between a stellar bulge
(highly asymmetric, dominating the potential from the center up to 3-4 kpc), a stellar disk (with possibly more than one
component), extending up to $\sim$15 kpc, and a disk of gas, lying approximately in the same plane of the stellar disk, and
mostly subleading in dynamical terms, which we include nonetheless for the sake of completeness.


In order to study the bulge, rather than relying on the spherical approximation common to many previous analysis, 
we adopt the approach described in~\cite{Pato:2015dua}, which takes into account a full three-dimensional
density distribution for stars in the Galactic bulge, solves the potential, and then finds the component within the disk, also allowing a precise
estimate of the lack of axisymmetry in that region.

The different morphologies for the Bulge and the Disc (separately) --collected and presented in 
\cite{Iocco2015} and then adopted by the same authors in \cite{Pato:2015dua}--
are inferred from observations of different population of stars
in different regions (see original references in \cite{Iocco2015, Pato:2015dua}), and therefore
fully empirical, three-dimensional, alternative descriptions of the stellar component of Bulge and Disk(s).
Following \cite{Iocco2015, Pato:2015dua}, we adopt --separately--
6 models of Bulge (labeled a,b,c,d,e,f) and 5 models of Disc (labeled I,J,K,L,M), which are then individually combined (one disk and one bulge at the time) 
thus obtaining a total of 30 combinations of Bulge plus Disc. To each of these possible stellar morphologies,
we add an observationally inferred morphology for the interstellar gas disk  taken from \cite{Ferriere:2007yq} from the inner 3 kpc and \cite{Ferriere:1998} above 3 kpc, instead of bracketing two possible alternatives, as done in \cite{Iocco2015, Pato:2015dua} given the subdominant contribution of the gas component to the RC.

For our fit we will thus have one (discrete) parameter to describe the uncertainty
related to the baryonic mass, namely the index of the baryonic morphology $\mathcal{M}_i$.
The normalization of each morphology, corresponding the mass of the Disc and mass of the Bulge however, also has its own uncertainty.
To take into account this uncertainty we normalize the morphology 
so that to agree with microlensing
optical depth measurements towards $(\ell,b)= (1.50^\circ, -2.68^\circ)$,
$\langle\tau\rangle = 2.17^{+0.47}_{-0.38} \times 10^{-6}$~\cite{Popowski:2004uv},
and local total stellar surface density   $\Sigma_{*} = 38 \pm 4\, {\rm M}_\odot /{\rm pc}^2$~\cite{Bovy:2013raa}. 
See again~\cite{Pato:2015dua} for more details. 
To take into account the uncertainty in $\langle\tau\rangle$ and  $\Sigma_*$ we will add them to the total
$\chi^2$ used to constrain the DM Halo and  vary them in the range $\pm 2\sigma$.
This is discussed in more details in section~\ref{sec:fitting}.

Finally, the morphologies depend on $R_0$. Thus, when changing the used value
of $R_0$ we self-consistently recalculate the morphology and its contribution to the RC.

\subsection{Local Standard of Rest}
Further uncertainty comes from the not well known peculiar motion of
the Solar system with respect to  the local standard of rest (LSR),
the system comoving along a circular orbit around the GC with a velocity
equal to the local RC velocity $V_0$.
Recent measurements find values $(U, V, W)_{\odot} = (11.10, 12.24, 7.25)$ km/s \cite{Schoenrich:2009bx},
where $(U, V, W)_{\odot}$ are, respectively, the velocity orthogonal to the circle of the orbit and  pointing outward the GC,
the velocity tangential to the circle, and orthogonal pointing in the $z$ direction.
The most relevant for the RC analysis is $V_{\odot}$ which in \cite{Schoenrich:2009bx} is found to be $V_{\odot}=12.24$ km/s,
but which has a quite larger scatter in the range $5 - 24$ km/s 
from different analyses in the literature \cite{Schoenrich:2009bx, BovyAPOGEE2012, Reid:2014boa}.
$V_{\odot}$ can be used together with the precise determination
of the total Solar system angular velocity $\Omega_{g, \odot} = 30.24\pm0.12\;{\rm km\,s^{-1} kpc^{-1}}$ \cite{blandhawthorn2016}
based on observations of the peculiar motion of the GC source Sagittarius A$^*$.
They are linked by 
\begin{equation}
\Omega_{g, \odot} = \frac{V_0 + V_{\odot}}{R_0}.
\label{eq:sun_angular_total}
\end{equation}
from which the local circular velocity $V_0$ can be derived once $R_0$ is also specified.
For \mbox{$R_0 = 8.0$ kpc,} Eq.~\ref{eq:sun_angular_total} gives  $V_0 = 230$ km/s,
which are commonly adopted values.
In the following we will use as free parameter $R_0$ which we will vary in the range $[7.5, 8.5]$ kpc.
We will instead fix $V_{\odot}=12.24$ km/s , since  the uncertainty in $V_{\odot}$
introduces a variation in $V_0$ similar or smaller than the one caused by $R_0$.
In practice, the uncertainty in $V_{\odot}$ can be taken into account by considering
a slightly more conservative range of variation for $R_0$.
Recently, the GRAVITY collaboration \cite{Abuter:2018drb} reported the very precise result $R_0=8.122\pm0.031$kpc.
If confirmed, this would essentially fix the value of $R_0$, so that in this case it would be convenient to consider 
explicitly $V_{\odot}$ as a parameter to vary in the analysis.

\subsection{Dark Matter distribution}
We parameterize the DM distribution as a spherically symmetric generalized NFW profile~\cite{Navarro:1996gj}
\begin{equation}
  \rho(r)=  \rho_s     \left( \frac{r}{R_s}\right)^{-\gamma}   \left(  1 + \frac{r}{R_s} \right)^{-3+\gamma}
\label{gNFW}  
\end{equation}
where $r$ is the spherical distance from the Galactic center (GC), $R_s$ the scale radius of the profile and
$\rho_s$ the scale density. The density behaves like $r^{-\gamma}$ toward the GC, and
the case $\gamma=1$ denoted the standard NFW profile.
The fit will thus have 3 parameters related to DM, $R_s$, $\rho_s$, and $\gamma$.

\begin{figure}
\centering
\includegraphics[width=0.8\columnwidth]{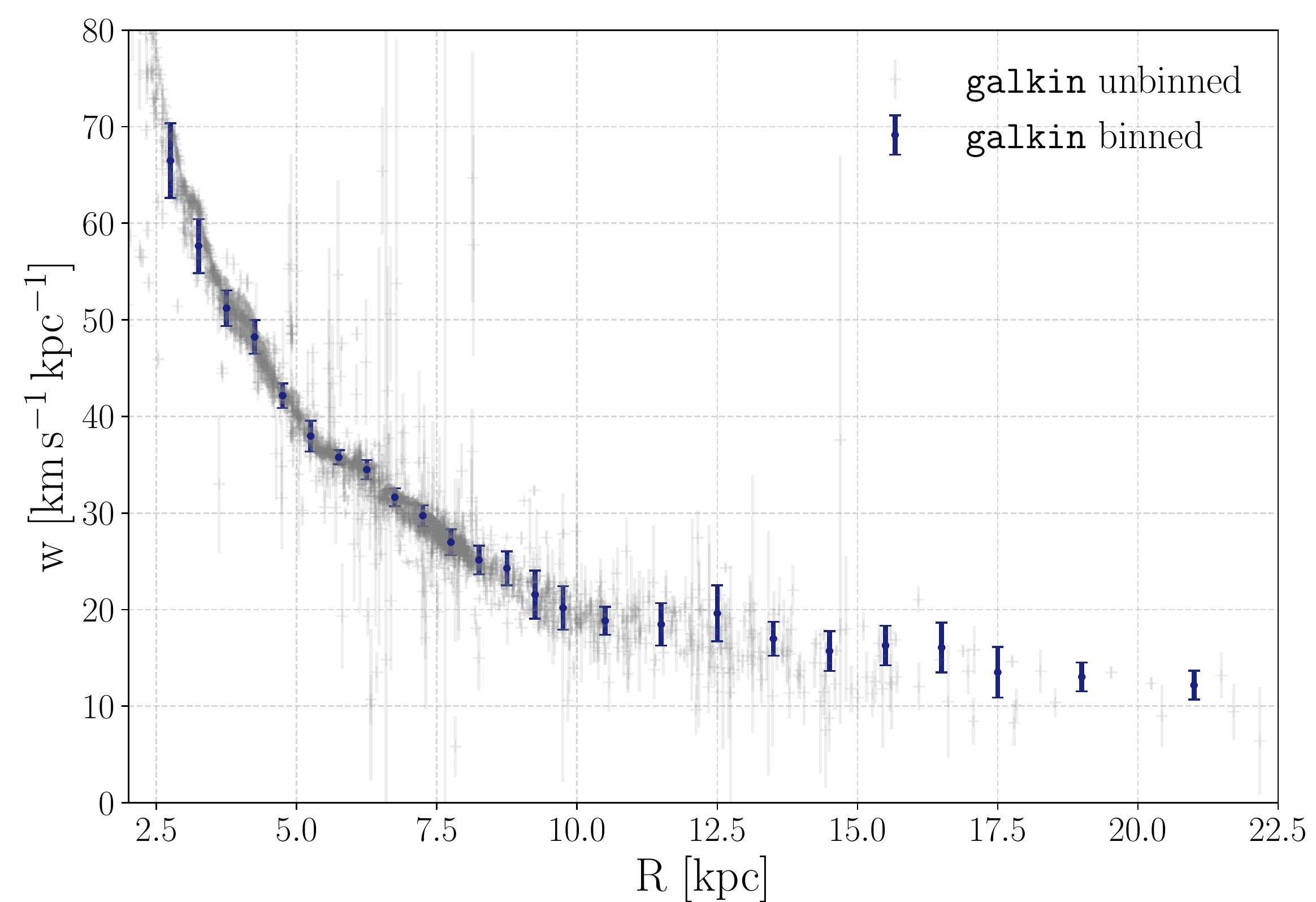}
\caption{Rotation curve of the Milky Way. {\tt Galkin} compilation and binned data for $R_0=8$ kpc, $V_0=230$ km/s, 
and $(U, V, W)_{\odot}$ from \cite{Schoenrich:2009bx}.}
\label{fig:w_vs_r}
\end{figure}

\section{Methodology}
\label{sec:method}

For our analysis, we adopt the angular velocity rotation curve $\omega(R)$ instead of the linear
velocity $V(R)$ in order to get rid of existing correlations between the uncertainty of the latter 
and that of the galactocentric distance $R$ \cite{Pato:2017yai}.

\subsection{Binning scheme}
\label{sec:binning}
The binning of the data of the observed RC is a key point, as we adopt binned data 
instead of an unbinned analysis such as that performed in e.g.\cite{Iocco2015,CatenaUllio2010}.
An unbinned analysis would exploit the full constraining power of the data
assuming that the underlying systematics uncertainties are under control.
Systematics of diverse nature are possible. For example,
single data points have sometimes very small errors, which, however, might not necessarily reflect
the true uncertainty of the RC which can have systematic contributions,
e.g. from peculiar motion of the gas tracers thus not following exactly
a circular orbit, and more in general from deviation from axial symmetry
of the motion of the tracers (see e.g. the Supplementary information of \cite{Iocco2015} and references therein).
Whereas the systematic errors
have been shown not to affect more general conclusions \cite{Iocco2015},
they are important in the details of the determination of the DM profile \cite{Pato:2015dua}.
Here we will thus employ a conservative view and use binned data.

We will hence
consider binned data and as error the dispersion of the data in the bin rather than a formal
weighted mean of the data.
Furthermore, we bin the data in $x=R/R_0$ rather than $R$ itself.
This because when varying $R_0$ the data `move' along the $R$ axis.
Binning in $x$ mitigates this problem, so that for different $R_0$
a given $x$ bin contains roughly the same unbinned data-points.
We start from $x=2.5/8$. Data below this value of $x$
are not considered in the fit in order to avoid the inner Galaxy
region for which there are significant deviations from
an axysimmetric motion of the tracers of the gravitational potential.
More in detail, we apply the following binning scheme:
\begin{itemize}
\item 15 bins from  $x=2.5/8$ to $x=10/8$   with a  step of $\Delta x=0.5/8$,  
\item 7 bins from    $x=10/8$ to $x=18/8$   with a  step of $\Delta x=1/8$,  
\item 2 bins from    $x=18/8$ to $x=22/8$   with a  step of $\Delta x=2/8$,  
\end{itemize}
for a total of 25 $x$ bins.
A bin of $\Delta x=0.5/8$ or smaller for $x<10/8$ is necessary to properly
follow the $\omega(x)$ curve  which is quite steep in this $x$ range.
At large $x$ the binning is larger also due to the scarcity of data.
In order to assign a given data-point to a bin we consider only the central value $x_i$
and neglect the error $\delta x_i$. 
Within each bin the center of the binned data-point $\bar{\omega}$ and its uncertainty $\sigma_{\bar{\omega}}$ are constructed as follow
\begin{equation}
\bar{\omega} =\frac{\sum_{i=1}^{N_{bin}}\omega_i/\sigma_{\omega_i}^2}{\sum_{i=1} ^{N_{bin}}1/\sigma_{\omega_i}^2},
\label{eq:binmean}
\end{equation}
\begin{equation}
\sigma_{\bar{\omega}}^2 =\frac{\sum_{i=1}^{N_{bin}}(\bar{\omega}-\omega_i)^2/\sigma_{\omega_i}^2}{\sum_{i=1} ^{N_{bin}}1/\sigma_{\omega_i}^2}  + \frac{N_{bin}}{\sum_{i=1} ^{N_{bin}}1/\sigma_{\omega_i}^2},
\label{eq:binsigma}
\end{equation}
i.e., $\bar{\omega}$ is just the weighted mean of the data in the bin.
$N_{bin}$ is the number of data points in the bin.
The uncertainty $\sigma_{\bar{\omega}}$ is composed of two terms,
the first is the weighted dispersion of the data,
the second term  gives the mean weighted error of the data, so that
the final error of the binned data-point will be larger than the latter.
The second term is however subdominant, i.e., the scatter in the data is much larger than the average error, typically 
by one order of magnitude of more, so the role of the latter is marginal,
 except for the last 2-3 bins, for which  the two errors are comparable. 
Finally, before applying the above procedure we take 
unbinned data-points above $x=10/8$ with a relative error of less than 10\%
and we increase artificially their uncertainty to 10\%.
This is to avoid that when  few data are present in a single bin
the final result gets dominated by a single data-point with very small error.
In practice, however, this affects only a single binned data-point, namely the 24th.
An example of the binned and unbinned RC for $R_0=8$ kpc, $V_0=230$ km/s, 
and $(U, V, W)_{\odot}$  from \cite{Schoenrich:2009bx} is given in Fig.~\ref{fig:w_vs_r}.

In the Appendix we will show the effect of using a different binning scheme to study the impact
on the final result.


\subsection{Fitting procedure}
\label{sec:fitting}

We include in the fit 7 parameters, i.e.,  $R_s$, $\rho_s$, $\gamma$, $R_0$, $\mathcal{M}_i$, $\langle\tau\rangle$ and $\Sigma_*$.
The number of parameters is still sufficiently small to use a grid scan rather than a Monte Carlo scan.
In the following we will thus just use a discrete grid. More precisely, we use 
50 values for $\rho_s$ linearly spaced in the range $[0.0,2.0]$ GeV/cm$^3$,
50 values for $R_s$ logarithmically spaced in the range $[5.0,100.0]$ kpc,
15 values of $\gamma$ linearly spaced in the range $[0.0, 1.5]$,
11 values of $R_0$ linearly spaced in the range $[7.5, 8.5]$,
and  30 morphologies $\mathcal{M}_i$.
For $\langle\tau\rangle$ and $\Sigma_*$ and we use 10 values each, linearly spaced in the
range $[-2\sigma,+2\sigma]$.

Having specified the above methodology to bin the data, we then compare them with the model using 
a simple $\chi^2$ statistics,
\begin{equation}
 \chi^2_{\rm RC}(R_s, \rho_s,  \gamma, R_0, \mathcal{M}_i) = \sum_j   \frac{\left(  \bar{\omega}_j - \bar{\omega}_j^{th}   \right)^2}{
          \sigma_{\bar{\omega}_j}^2   }   + \chi^2_{\langle\tau\rangle}(\langle\tau\rangle) + \chi^2_{\Sigma_*}(\Sigma_*)
 \label{eq:chi2RC}         
\end{equation}
and we evaluate $\chi^2_{\rm RC}$ over the grid defined above.
$\bar{\omega}_j^{th}$ is the model prediction depending on $R_s, \rho_s,  \gamma, R_0, \mathcal{M}_i$, $\langle\tau\rangle$ and $\Sigma_*$
and is given by $\bar{\omega}_j^{th} =\sqrt{\left(\bar{\omega}_j^{b}(R_0, \mathcal{M}_i,\langle\tau\rangle, \Sigma_*)\right)^2 + \left(\bar{\omega}_j^{DM}(R_s, \rho_s,  \gamma)\right)^2}$,
i.e., by the sum of the DM and baryonic contribution.
For  $\chi^2_{\langle\tau\rangle}$  and $\chi^2_{\Sigma_*}$ we use the expressions $\chi^2_{\langle\tau\rangle}=(\langle\tau\rangle-2.17)^2/0.42^2$
and  $\chi^2_{\Sigma_*}=(\Sigma_*-38)^2/4^2$ from section~\ref{sec:bmorphology}, where the error on $\tau$ has been made symmetric
for simplicity.
We verified that including $\langle\tau\rangle$ and $\Sigma_*$ in the $\chi^2$ does not crucially affect the analysis.
Keeping $\langle\tau\rangle$ and $\Sigma_*$ fixed to their central values only reduces slightly the error
in the determination of the other parameters of the analysis. This is likely due to the
fact that the bulk of the uncertainty from the baryonic morphology is already taken into
account considering the 30 different models  $\mathcal{M}_i$. Nonetheless,
for consistency of the analysis and for a more robust error determination, we include $\langle\tau\rangle$ and $\Sigma_*$
in the overall $\chi^2$. Furthermore, again for the above reason, just 10 grid values of $\langle\tau\rangle$ and $\Sigma_*$ are 
already enough to properly include the effect of their uncertainty on the analysis.

Another point to mention is that, formally the above definition is not fully self-consistent since the data change when $R_0$
is changed and we explore changes in $R_0$ of the order of 10\%, which introduces
changes in the data of the same order. This issue is unavoidable as soon as binned data are used.
Nonetheless, as long as the induced variations in the data are smooth
as function of $R_0$, as it is the case, the data variation can be thought as being
reabsorbed into a redefinition of the model, so that the use of Eq.~\ref{eq:chi2RC}
should be approximately valid. 
Another minor inconsistency is given by the fact that once the set of 7 
parameters is specified, the full rotation curve $V(R)$ is also specified
and so is $V_0 \equiv V(R_0)$. The relation $V_0=V(R_0)= R_0 \sqrt{\omega_b^2(R_0) + \omega_{DM}^2(R_0)}$ should thus be enforced
and used to remove one parameter. 
This, in practice, is not a big issue, since the fit will
automatically prefer the region where this relation is satisfied.
Furthermore, this extra freedom is, in practice, equivalent to not
strictly assume  Eq.~\ref{eq:sun_angular_total} linking $V_0$ and $R_0$ but leaving
some freedom in their relation to be constrained by the fit,
which is a conservative choice.

\begin{figure}[t]
\centering
\includegraphics[width=1.0\columnwidth]{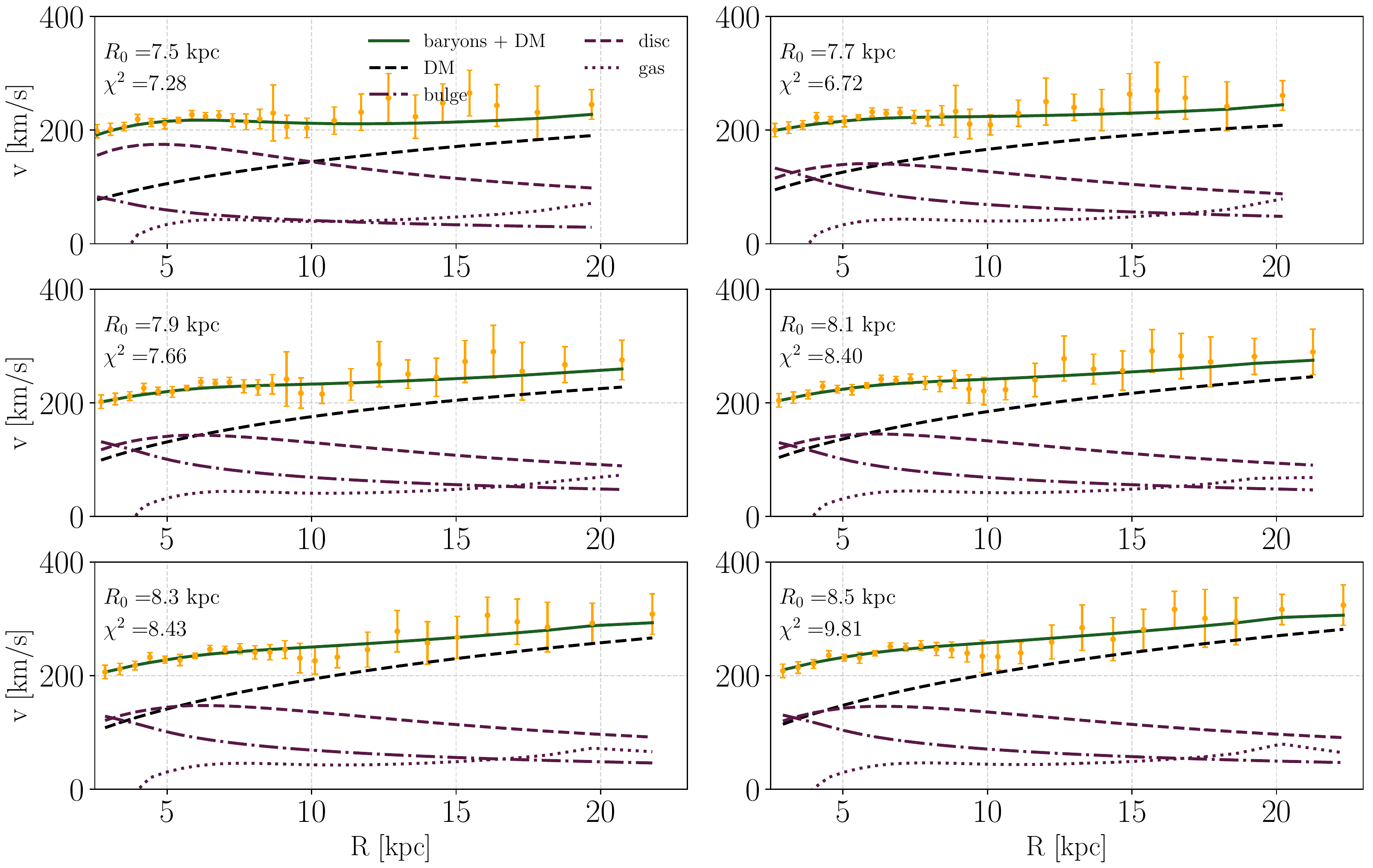}
\caption{Rotation Curve fit results for $\gamma=1$ and different values of $R_0$.
The various curves represent the best-fit contribution to the RC from the Bulge, Disc, gas,  DM, and the total as labeled in the plot
(see text for more details).}
\label{fig:binning_BF_xy}
\end{figure}

We use a fully frequentist framework to derive constraints in sub-spaces
of  the 7 full dimensional space. Specifically we employ the commonly used method
of profiling \cite{Rolke:2004mj}. For example, when we build two-dimensional $\chi^2$ in two
given parameters, for each 2-d grid point, we take the minimum $\chi^2$ over the remaining 5 parameters.
We proceed similarly when building 1-d, 2-d, or 3-d profiled $\chi^2$.

We will show in the following, as an example, the constraints on the parameters from the above $\chi^2$.
Nonetheless, these constraints are not necessarily the optimal ones. For example,
the use of a flat prior on $R_0$ in the range $[7.5, 8.5]$ kpc is perhaps too conservative
and, instead, more stringent priors could be used, as for example the Gaussian prior
$R_0=8.2\pm 0.1$ kpc based on Ref.~\cite{blandhawthorn2016}. 
Similar considerations  apply to $\gamma$ or to the other parameters.
This extra information can be easily included starting from the tables we provide. The main goal of this analysis
is to provide results in a general form
such that they can be used by the community together with complementary information, 
with the aim to simplify the use of a thorough data--driven approach on astrophysical uncertainties
to analysis including direct and indirect DM searches, as well as collider probes.

\begin{figure}[t]
\centering
\includegraphics[width=0.45\columnwidth]{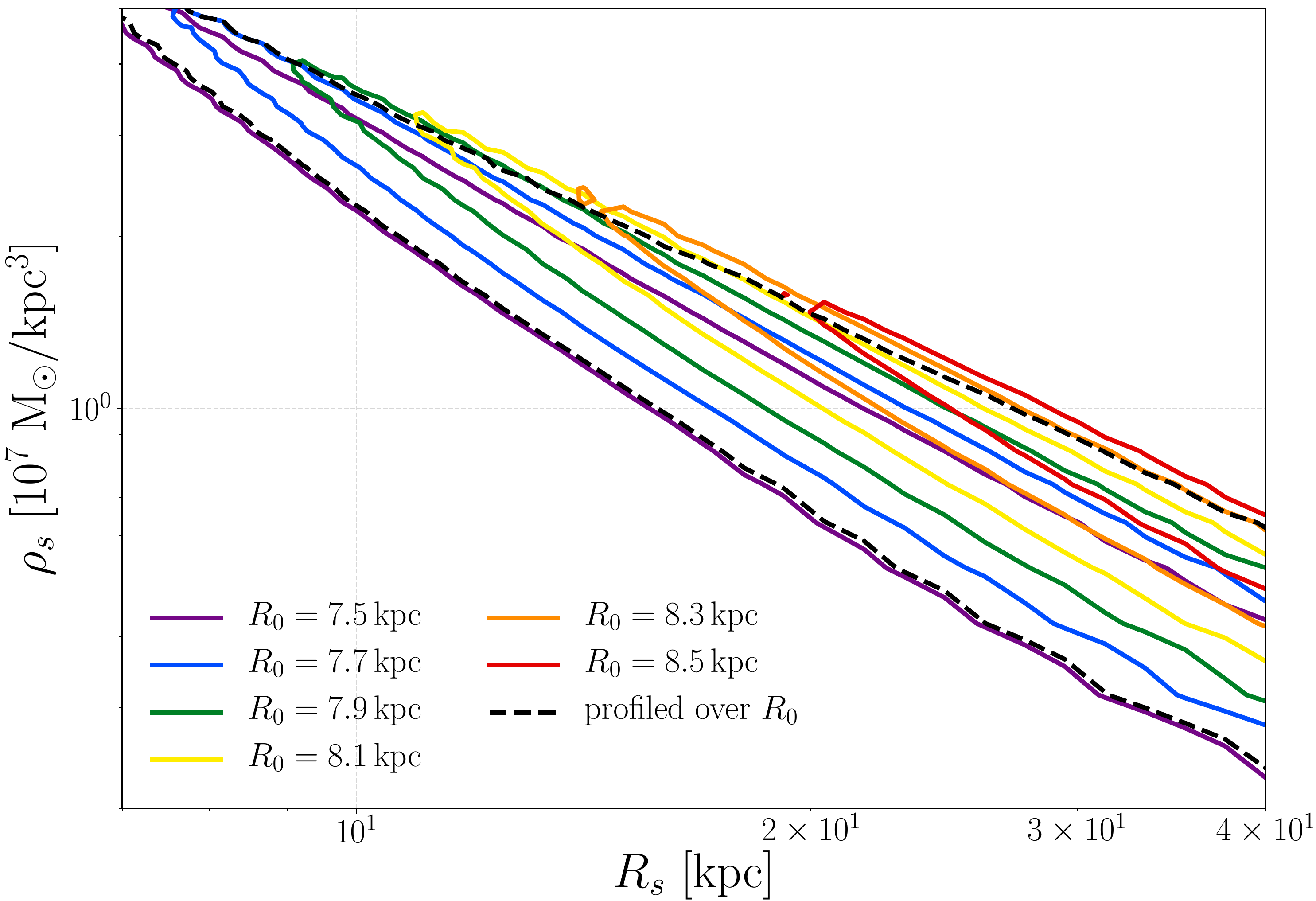}
\includegraphics[width=0.45\columnwidth]{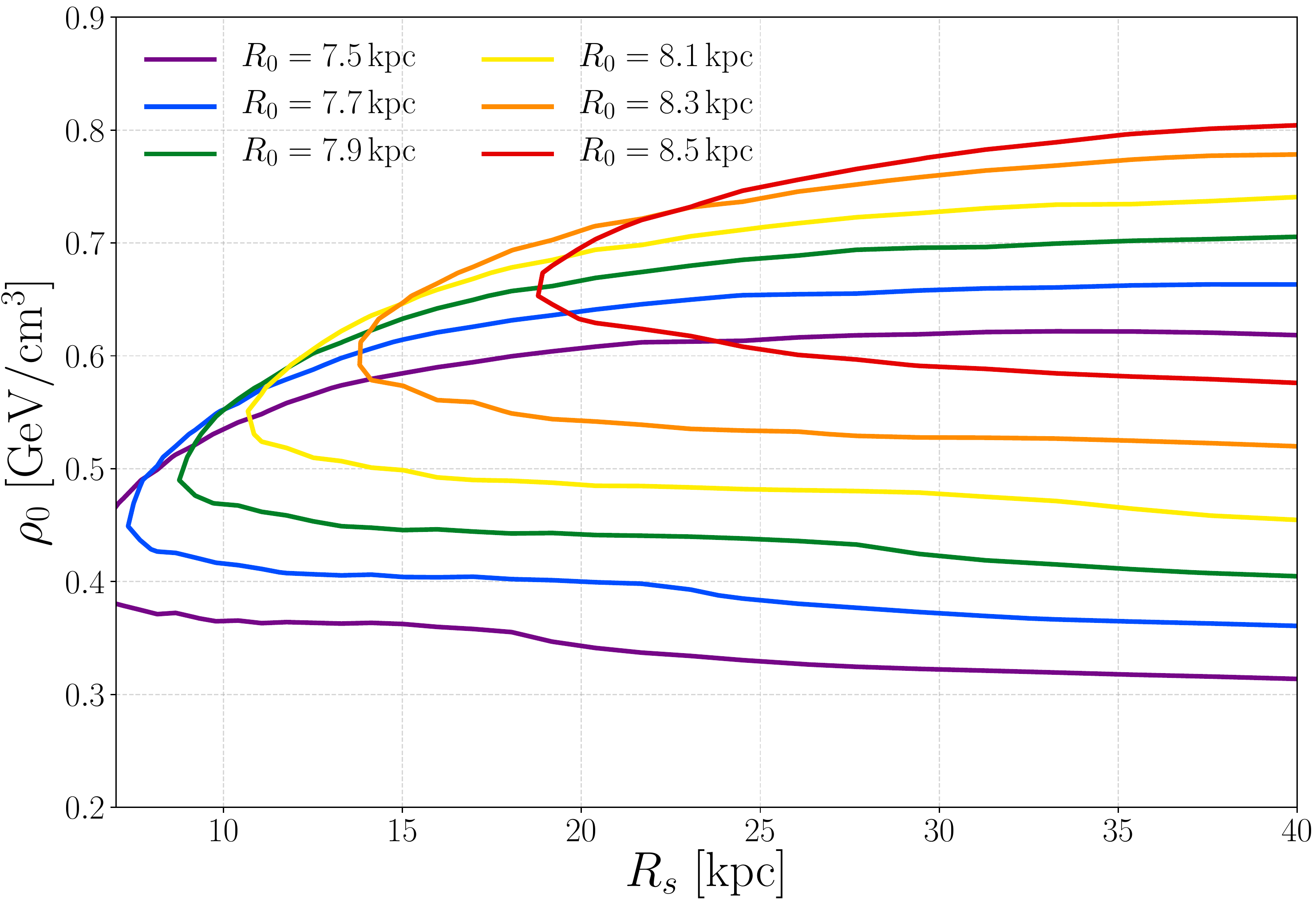}
\includegraphics[width=0.45\columnwidth]{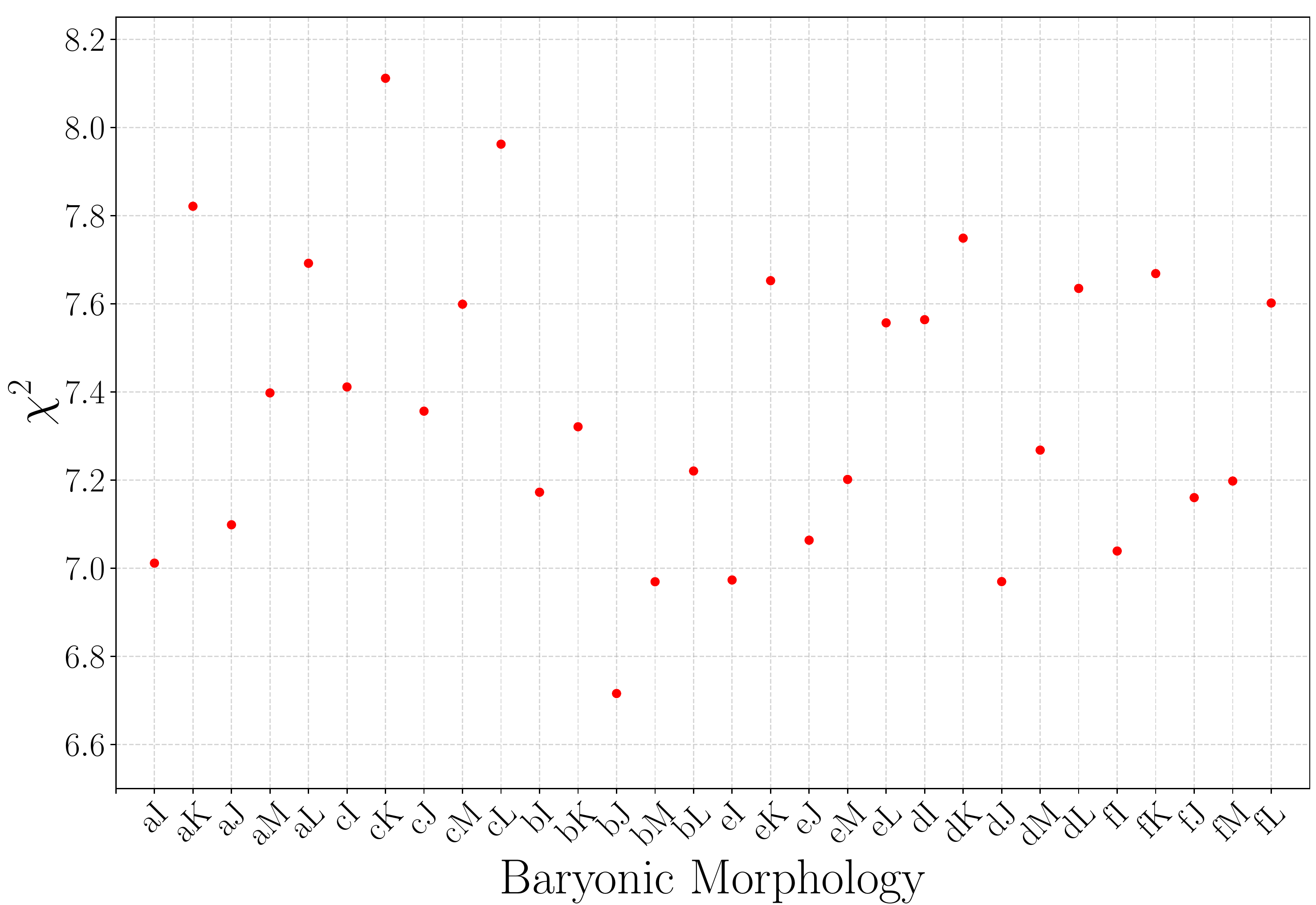}
\includegraphics[width=0.45\columnwidth]{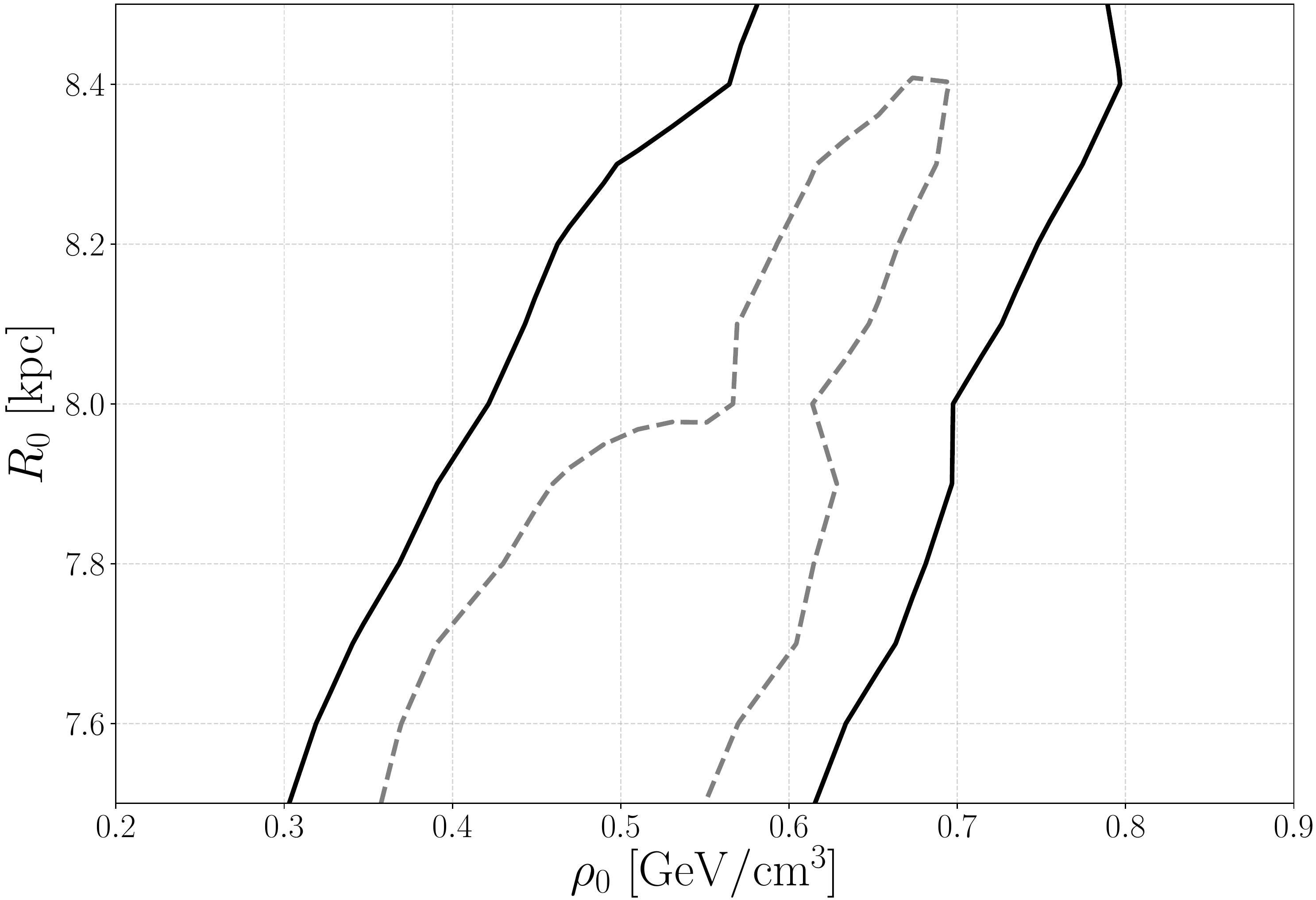}
\caption{Upper-left panel: 2$\sigma$ contours in the $(R_s,  \rho_s)$ plane for fixed $\gamma=1$
and for  various values of $R_0$ and profiled over $\mathcal{M}_i$, $\langle\tau\rangle$ and $\Sigma_*$ as well as the final
contour profiled over $R_0$, $\mathcal{M}_i$, $\langle\tau\rangle$ and $\Sigma_*$.
Upper-right: same as upper-left but in the $(R_s, \rho_0)$ plane.
Lower-left: 1-d profile $\chi^2$ of the baryonic morphology $\mathcal{M}_i$ profiled over $R_0$, $R_s$, $\rho_s$, $\langle\tau\rangle$ and $\Sigma_*$.
Lower-right: 1-2 $\sigma$ contours in the profile $\chi^2$ in the $(\rho_0, R_0)$ plane.}
\label{fig:contours_super_allR0}
\end{figure}

\section{Results }
\label{sec:results}

\subsection{NFW $\gamma=1$ case}
\label{resultsNFW}

In Fig.~\ref{fig:binning_BF_xy}, we show the result of the fit  for $\gamma=1$, i.e. the canonical NFW case,
and different values of $R_0$.
As explained above we effectively fit the $\omega(x)$ data, but
Fig.~\ref{fig:binning_BF_xy} shows the fit results and data-points
in the $V(R)$ plane, which gives a more familiar representation. 
It can be also seen that since the data-points are at fixed $x$ values,
when shown as function of $R$ they move along the $R$ axis for different $R_0$ values.
The various curves show the best-fit contribution to the RC from the different baryonic components
(Bulge, Disc and gas) and DM, as well as the total.
The gas component, as explained previously, is the same since it is not varied
in the fit. The morphology instead can be different for the various cases.
Speficically, Bulge {\it b}  is the preferred bulge morphology for all cases, except $R_0$= 7.5 kpc which is bulge {\it e}. 
All $R_0$ cases prefer Disc {\it j}, except for $R_0$ = 7.5 kpc which prefers Disc {\it i}.
The plot also lists the best-fit $\chi^2$ which are in the range 7-10 for 25 data-points and 7 fitting parameters
for a reduced $\chi^2$ $\simeq 8/(25-7)\simeq0.5$.
The value is slightly low and indicates that the binning procedure somehow overestimates
the errors. Nonetheless, since this will give conservative results, we prefer not to modify our procedure.

Further results of the fit are shown in Fig.~\ref{fig:contours_super_allR0}.
The upper-left panel shows 2-$\sigma$  ($\Delta \chi^2=6.18$ from the minimum for two degrees of freedom) contours
in the $\rho_s - R_s$ plane for fixed values of $R_0$ and profiled over $\mathcal{M}_i$, $\langle\tau\rangle$ and $\Sigma_*$,
as well as for the case profiled over both $R_0$ and $\mathcal{M}_i$, $\langle\tau\rangle$ and $\Sigma_*$ (black contour),
and shows a strong degeneracy between $\rho_s$ and $R_s$.
Upper-right panel shows, however, that when we visualize the results in the $\rho_0 - R_s$
plane, where $\rho_0$ is the local DM density (which  is a derived parameter in our framework)
the degeneracy basically disappears.
To generate this plot we used as independent parameter to build the grid $\rho_0$, rather than 
$\rho_s$  (from Eq.~\ref{gNFW} the two are related by  $\rho_0=  \rho_s     \left( R_0/R_s \right)^{-\gamma}   \left(  1 + R_0/R_s \right)^{-3+\gamma}$ ).
Interestingly the plot also shows that the constraints on $\rho_0$ strongly depend on $R_0$.
This is best seen in the lower-right panel where 1-$\sigma$ ($\Delta \chi^2=2.30$ from the minimum) 
and 2-$\sigma$ contours in the $R_0-\rho_0$ plane, profiled
over the remaining parameters, are shown.
The plot shows that the analysis is  sensitive to $R_0$, although not strongly.
This is reasonable, since $R_0$ is better constrained by different types of analysis
than the RC ones (see~\cite{Malkin:2013ac} for a list of works on the determination of $R_0$). 
In the absence of strong priors on $R_0$, $\rho_0$ values between
0.3 and 0.8 GeV/cm$^3$ at 2-$\sigma$ are allowed, which is in agreement with the conservative
estimate provided in \cite{Salucci2010}. Interestingly, the combination $\rho_0=0.4$ GeV/cm$^3$,
$R_0=8$ kpc, often used in the literature,  is in tension at 2$\sigma$ level with the fit result.
The previous ``standard'' used until recently,  $\rho_0=0.3$ GeV/cm$^3$,
$R_0=8.5$ kpc has a $\chi^2$ of 53.6 (in the profiled $\rho_0$, $R_0$ plane) and  it's excluded at more than 4~$\sigma$ confidence level.

\begin{figure}[t]
\centering
\includegraphics[width=0.45\columnwidth]{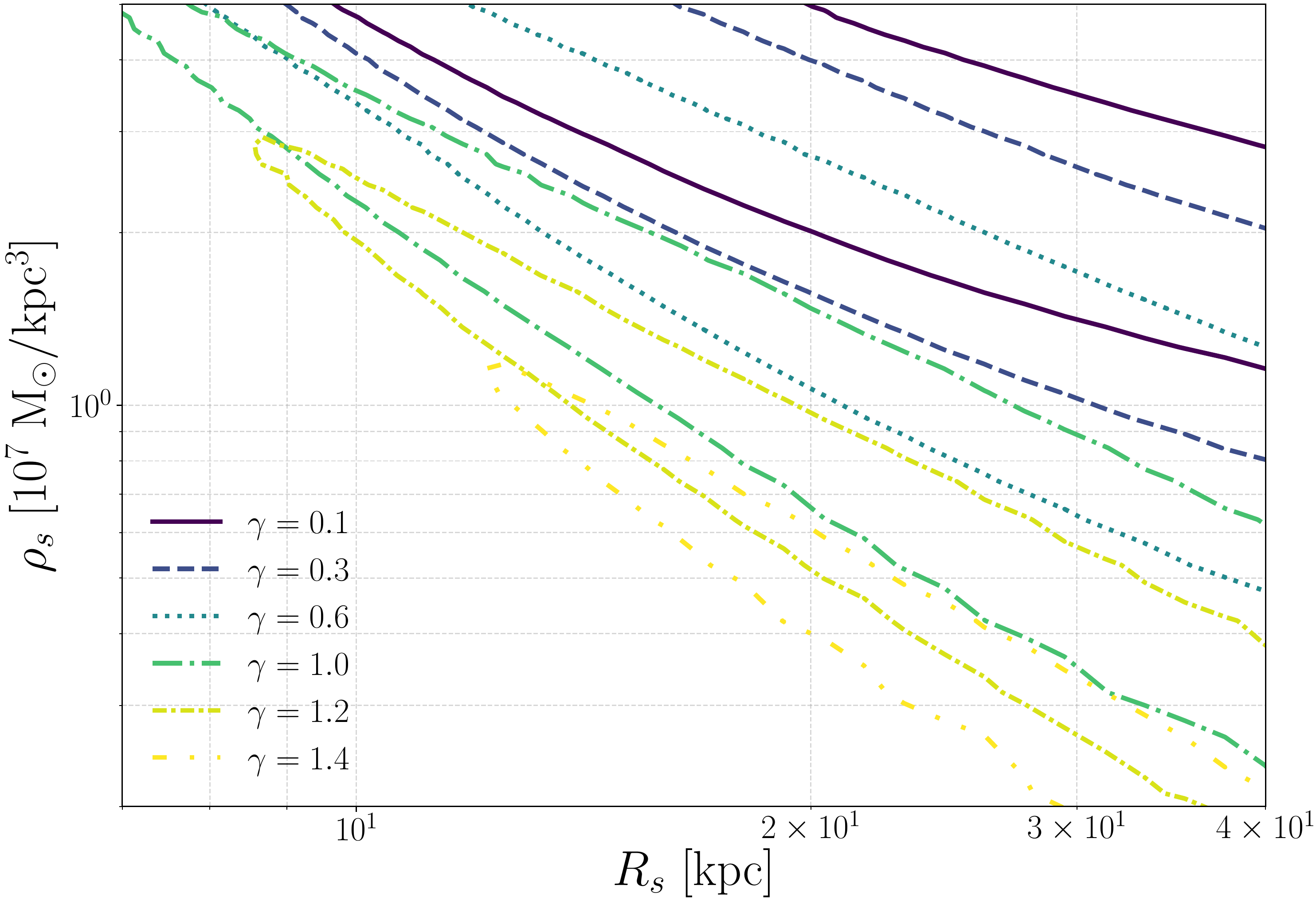}
\includegraphics[width=0.45\columnwidth]{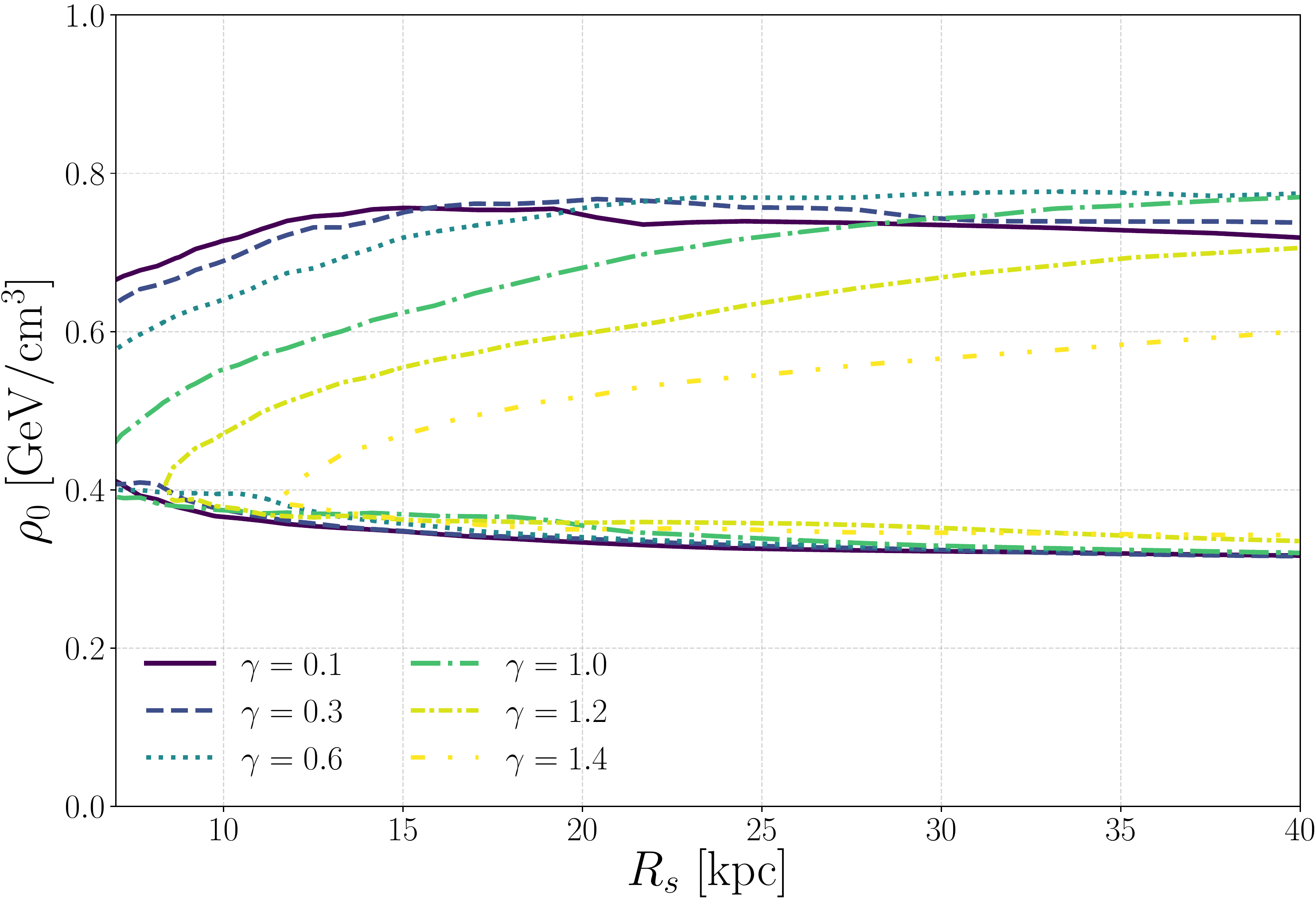}
\includegraphics[width=0.45\columnwidth]{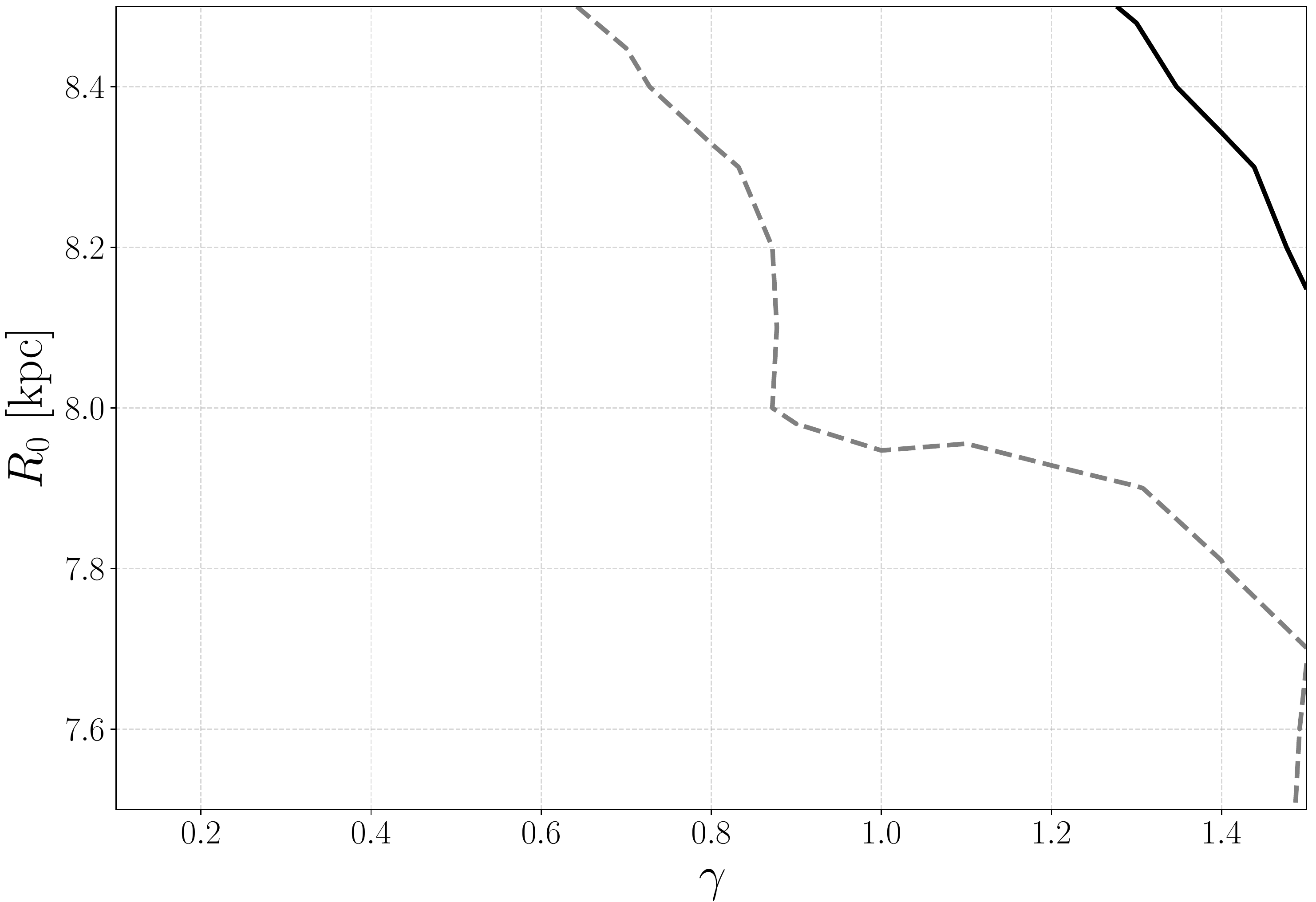}
\includegraphics[width=0.45\columnwidth]{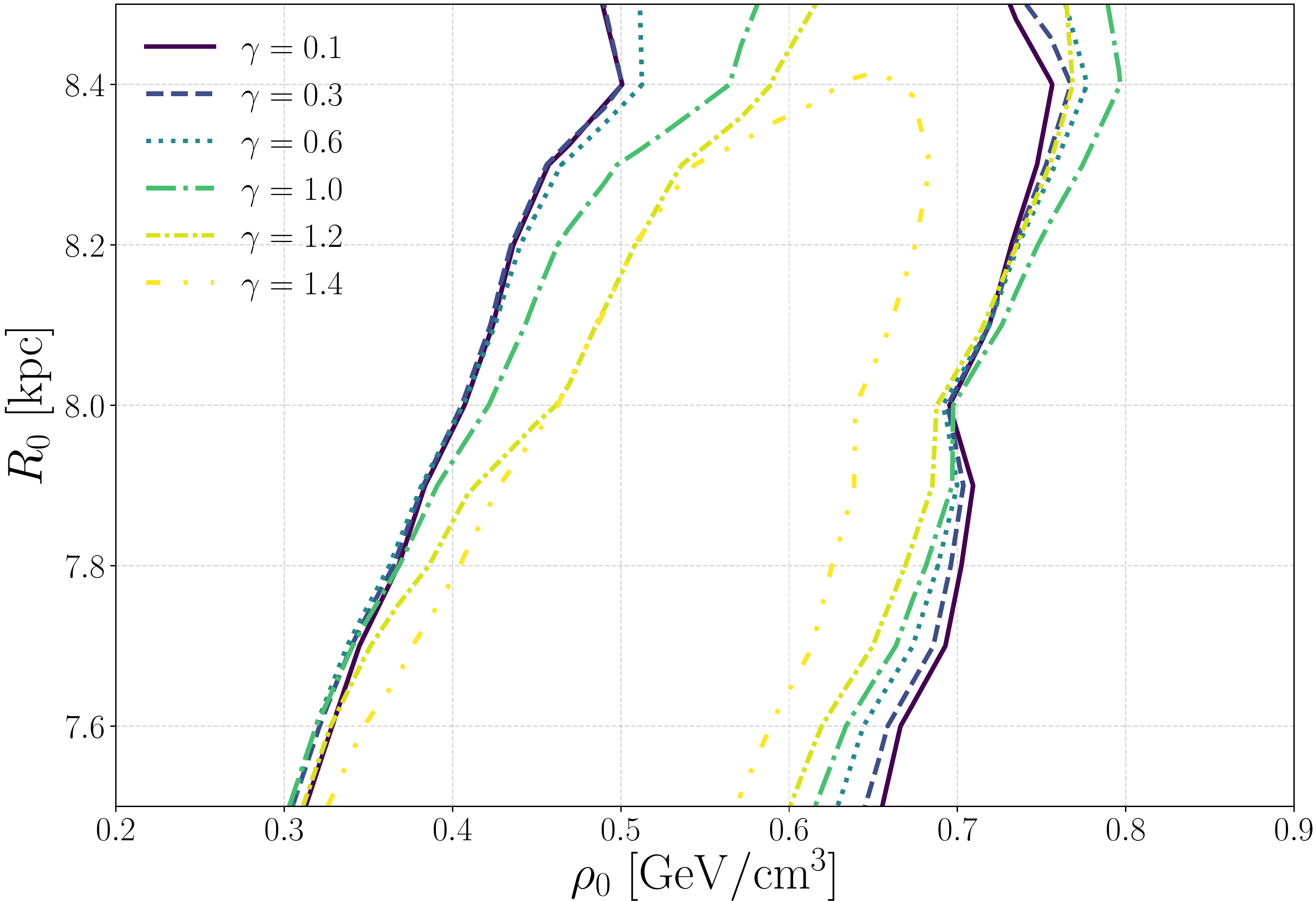}
\caption{Upper-left panel: 2$\sigma$ contours in the $(R_s,  \rho_s)$ plane for
 various fixed values of $\gamma$ profiled over $\mathcal{M}_i$, $\langle\tau\rangle$, $\Sigma_*$ and $R_0$.
Upper-right: same as upper-left but in the $(R_s, \rho_0)$ plane. 
Lower-left: 1-2 $\sigma$ contours of the $\chi^2$  in the $\gamma, R_0$ plane profiled over $\mathcal{M}_i$, $\langle\tau\rangle$, $\Sigma_*$, $R_s$, and  $\rho_s$.
Lower-right: 2-$\sigma$ contours of the profile $\chi^2$ in the $(\rho_0, R_0)$ plane for different fixed values of $\gamma$.}
\label{fig:contour_Rsrhos_Rsrho0_gamma}
\end{figure}

\begin{figure}[t]
\centering
\includegraphics[width=0.95\columnwidth]{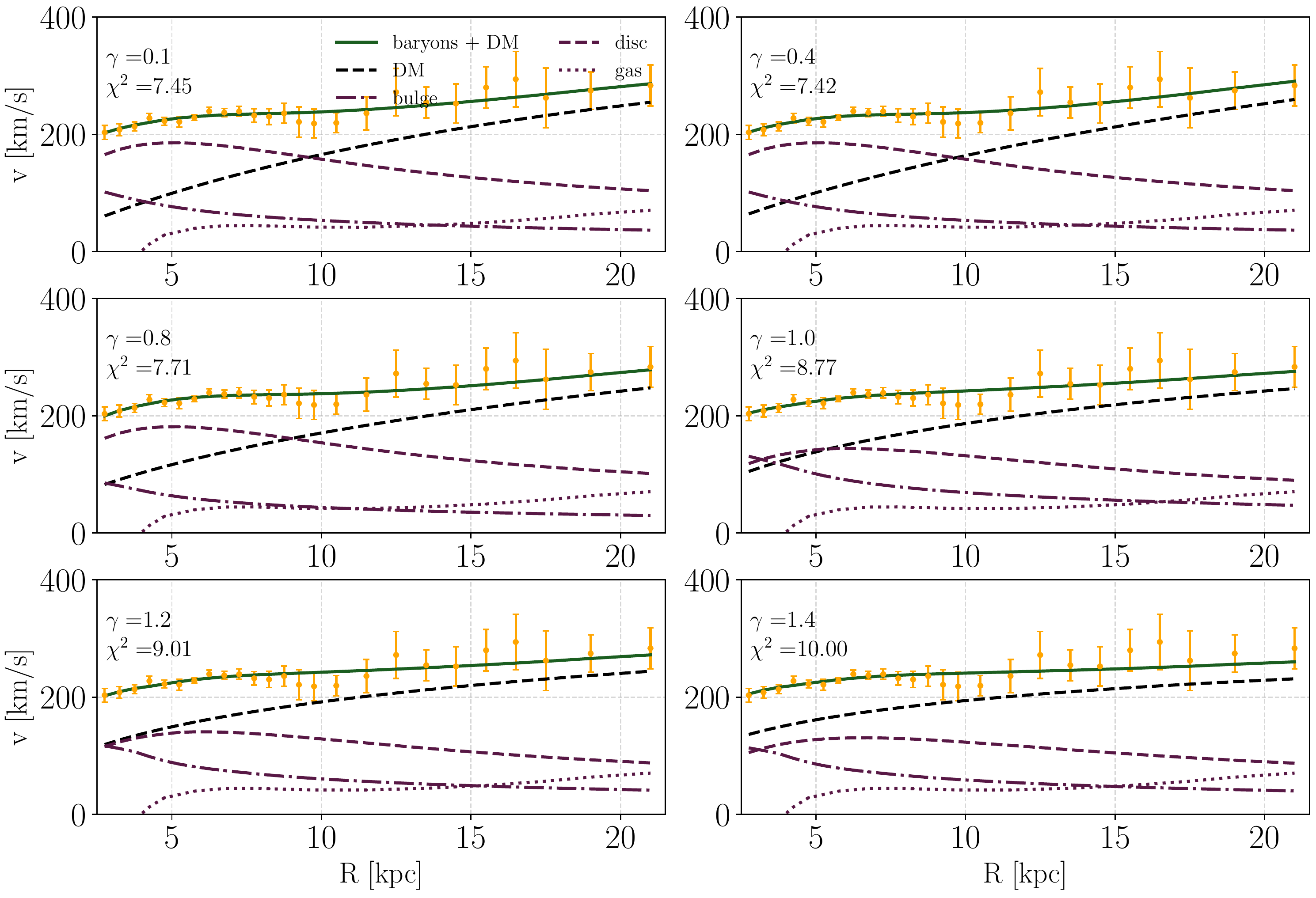}
\caption{Rotation Curve fit results for $R_0=8.0$ kpc and for different values of  $\gamma$ as labeled in the text.
The various curves represent the best-fit contribution to the RC from the Bulge, Disc, gas,  DM, and the total similarly to what shown in Fig.~\ref{fig:binning_BF_xy}.}
\label{fig:RCgamma}
\end{figure}

Finally, the lower-left panel displays the 1-d $\chi^2$ plot of the morphology $\mathcal{M}_i$
profiled over the remaining parameters, and it shows that no single morphology is
preferred by the analysis but all of them give similarly good best-fit $\chi^2$
at the level of 1-$\sigma$ or slightly more.
This confirms the importance of considering different morphologies in order not to bias the
final results. Comparing results obtained fixing the morphology to a single one
we find that the systematic effect on $\rho_0$ is around $\pm 0.1$ GeV/cm$^3$.
In particular, the two morphologies which are found to give results which differ the most
are the ones presenting a model of single disc vs the ones with a double disc (see \cite{Iocco2015, Pato:2015dua} for more details on the morphologies).
An example of fit with a fixed morphology is discussed in the Appendix.
The structure of the degeneracy in the  $(\rho_0, R_0)$ plane is, instead, unchanged for each
single morphology, i.e., the slope of the degeneracy remains the same and no significant sensitivity to  $R_0$ is present.

\subsection{Results as function of $\gamma$}
\label{resultsgamma}

In Fig.~\ref{fig:contour_Rsrhos_Rsrho0_gamma} we show
the analogous of Fig.~\ref{fig:contours_super_allR0} for the case in
which $\gamma$ is varied.
It can be seen that the results are similar, except for the fact that
when large values of $\gamma$ are used ($>1$), the largest values
of $\rho_0$ (in the range $0.6-0.8$ GeV/cm$^3$) are disfavored.
Also, in general, no constraints can be inferred on $\gamma$.
Including the uncertainty on $\gamma$, the couple  \mbox{$\rho_0=0.3$ GeV/cm$^3$},
$R_0=8.5$ kpc has now a $\chi^2$ of 42.3 (vs 53.6 when $\gamma$ is fixed to 1) which is still  excluded at more than 4~$\sigma$ confidence level.

The best-fit RCs are shown in Fig.~\ref{fig:RCgamma}
for a fixed value  of $R_0=8.0$ kpc and for different values of  $\gamma$.
It can be seen that despite strong differences in the RC contribution from DM in the cases $\gamma=0$ (i.e., cored profile)
and $\gamma=1$ (i.e., cuspy profile), an equally good fit can be achieved in both cases to the measured RC.
The main reason of this result is the degeneracy with the morphology. The uncertainties in the bulge and disc mass
and morphology are large enough that can compensate in the two cases the large change from the DM contribution.
The disc, in particular, seems to play a dominant role in this degeneracy, while the contribution from the bulge is slightly less prominent.
This also means that in the future a more precise determination of the bulge and disc mass
and morphology should be able to break this degeneracy and allow a reliable
determination of the inner slope $\gamma$.
A similar conclusion was reached by the work in \cite{Iocco:2016itg}, using a different
analysis involving only observations within $\sim 2$ kpc from the Galactic Center.

\subsection{Comparison with Other Results}
\label{sec:comparison}

As seen in the above sections, a general result of our analysis is that the single parameters
are only weakly constrained by the fit (even for the case of fixed NFW profile, i.e., without varying  $\gamma$).
For example, at $2 \sigma $ C.L. $\rho_0$ lies in the range
0.3-0.8 GeV/cm$^3$, thus with an error of \mbox{$\sim 0.25$ GeV/cm$^3$},  while
$R_0$ is simply unconstrained by the analysis with respect to the prior range 7.5-8.5 kpc. 
What  robustly constrained by the fit is, instead,
the degeneracy and correlation among the parameters, like, noticeably, the one between
$R_0$ and $\rho_0$. 
This is somewhat at odd with similar analyses performed in the past, which, typically,
tend to find very small errors and strong constraints on the parameters.
We attribute this difference to three main effects. First, the accurate statistical treatment,
which explores and maps in details the degeneracies among the parameters.
This is important, since if strong degeneracies are present, as in this case,
and they  are not well characterized, the error on the single parameters will be underestimated.
Second, as detailed in Sec.~\ref{sec:data}, we don't use stellar-tracer data up to $\sim$100 kpc,
sometimes adopted in other analysis. The DM potential is the dominant component in the range 20-100 kpc, 
so these data could actually provide an important contribution in reducing the DM halo parameters errors, 
although this comes at the cost of adding further assumptions.
Third, we use binned data, with an uncertainty estimated from the spread
of the datapoints in the bin.
This was already found to be an important point in \cite{Pato:2015dua}
which shares the same dataset and similar methods as the present analysis. 
More precisely, when using the unbinned analysis, in \cite{Pato:2015dua} the authors report   
$\rho_0  =  0.420^{+0.021}_{-0.018} (2\sigma) \pm 0.025$ GeV/cm$^3$
where the first error is statistical and the second comes from the baryonic morphology uncertainty. 
On the other hand, a test with binned data
gives errors a factor of $\sim 5$ larger, thus more compatible with the analysis performed here.
Other analyses giving small errors, like in \cite{CatenaUllio2010} which reports $\rho_0=0.389\pm0.025$ GeV/cm$^3$
also use unbinned data.
A noticeable exception is the work in \cite{Nesti:2013uwa}, where a binned analysis is performed, 
with bin errors estimated in a similar way as in this work. We thus expect 
an uncertainty similar to the one of our analysis. The authors indeed find \mbox{$\rho_0= 0.471^{+0.048}_{-0.061}$ GeV/cm$^3$},
which has an uncertainty larger than  \cite{Pato:2015dua,CatenaUllio2010}   but still smaller than our analysis.
This  is likely related to the larger dataset used in \cite{Nesti:2013uwa} which includes
stellar velocity dispersion measurements at $R>20$ kpc, or to  the simplified procedure used to estimate
the errors, as explained in \cite{Nesti:2013uwa}.
Finally, in \cite{Karukes2019}, the authors perform  a binned analysis using, for the inner Galaxy, the \textit{galkin} dataset also employed here, in combination, for the outer Galaxy, with  stellar tracers up to 100 kpc from \cite{Huang2016}.
The final uncertainties derived  there are thus smaller.
A further difference is that in \cite{Karukes2019} a fixed value of $R_0=8.34$ kpc is used, whereas one of the goals of this present analysis is indeed to estimate the very impact of the uncertainties on $R_0$ --which we therefore vary as discussed in the previous sections-- on the determination of the DM distribution.

Several other analysis exist in the literature, such as e.g.
\cite{Huang2016, McMillan2011, McMillan2017, DeBoer2009, Xue2008},
to which we address the reader for a sample of the wide range of methodologies, datasets and assumptions
employed.

\section{Implications for Direct and Indirect Dark Matter Searches }
\label{sec:indirect}

In this section, we provide examples of how to use the results derived above.
In particular we consider the example of the Galactic Center $\mathcal{J}$-factor uncertainty
for indirect DM searches and the uncertainties in direct DM searches.

\subsection{Galactic Center $\mathcal{J}$ factor}

\begin{figure}[t]
\centering
\includegraphics[width=0.45\columnwidth]{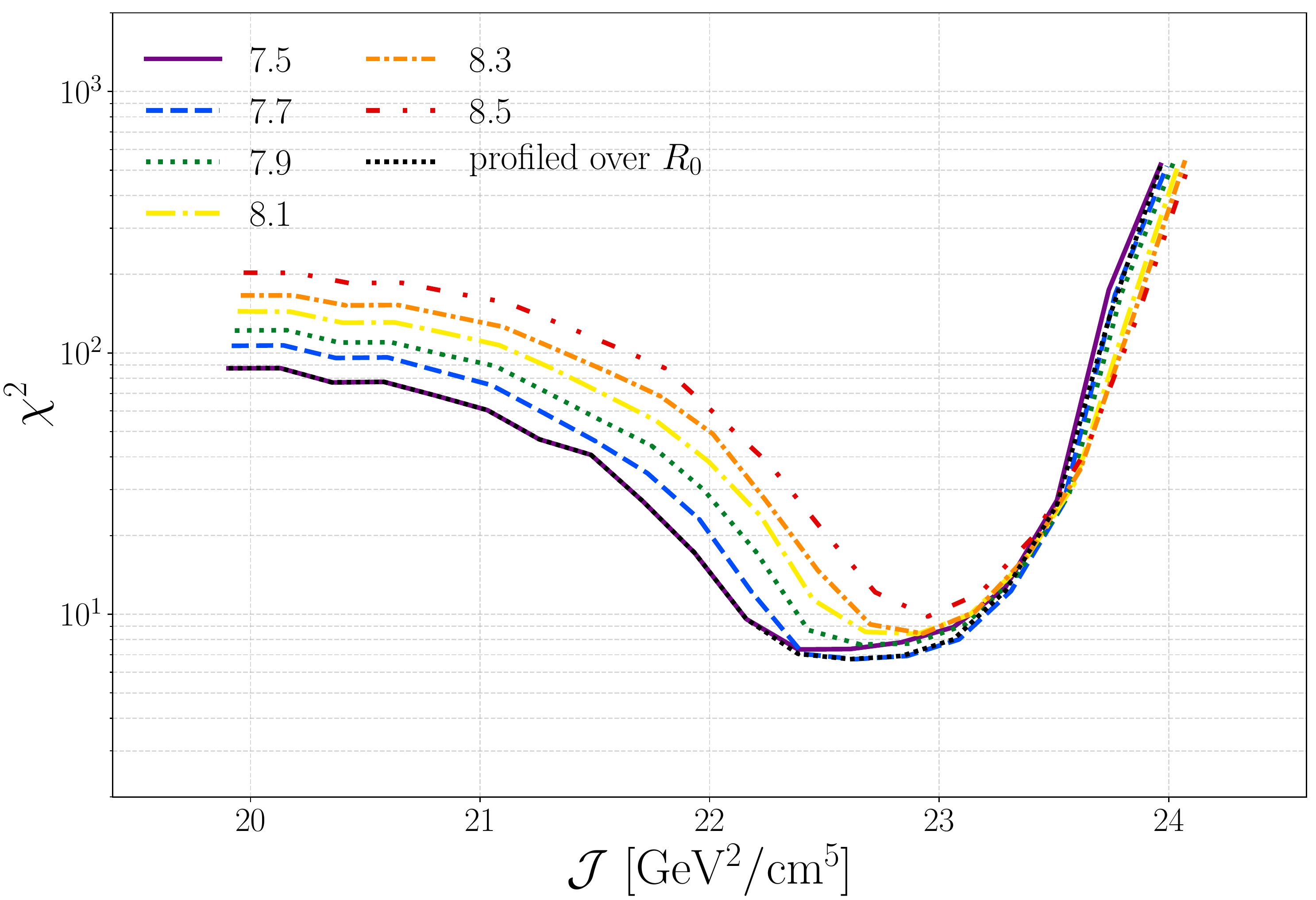}
\includegraphics[width=0.45\columnwidth]{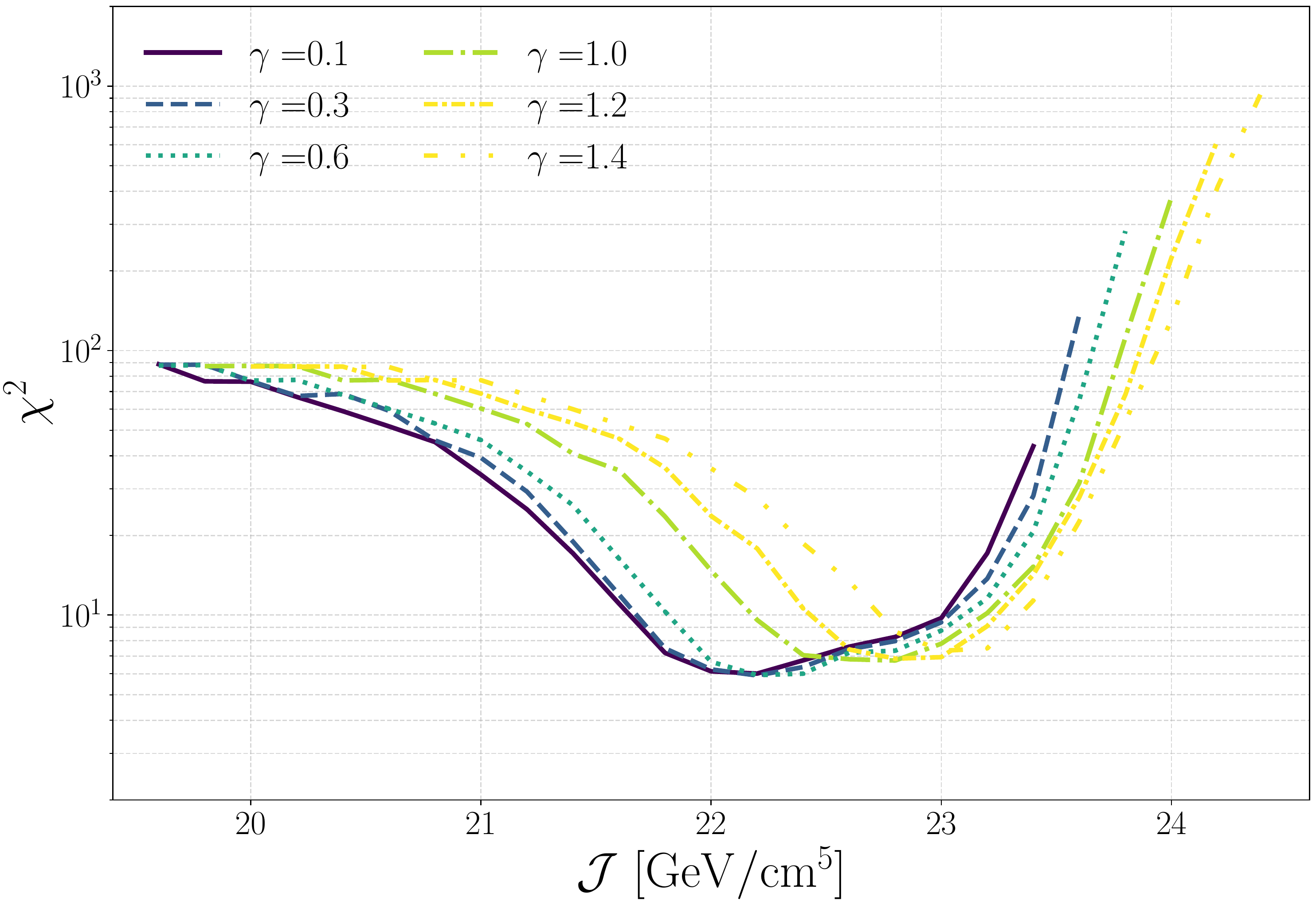}
\caption{Left panel: $\chi^2$ profile for the GCE $\mathcal{J}$-factor for fixed $\gamma=1$. 
The colour lines correspond to profiles over $\mathcal{M}_i$, $\langle\tau\rangle$, $\Sigma_*$, $R_s$ and $\rho_s$ for different $R_0$. 
The black-dashed line is the $\chi^2$ profiled furthermore over $R_0$.
Right panel: Colour lines correspond to the profiles over $\mathcal{M}_i$, $\langle\tau\rangle$, $\Sigma_*$, $R_s$, $\rho_s$  and $R_0$ for different $\gamma$. }
\label{fig:jfactor}
\end{figure}

An immediate application of the above analysis is the derivation of the GC $\mathcal{J}$-factor
and its uncertainty.  The GC $\mathcal{J}$-factor is given by
\begin{equation}
   \mathcal{J} =   \int_{\Delta\Omega}  \!\!\!\! d\Omega   \int_{\rm l.o.s.}   \!\!\!\! ds \, \rho^2(r(s,\psi))
\end{equation}
where $\psi$ is the angle from the GC, $s$ a coordinate along the line of sight, and $\Delta\Omega$
the region of interest over which the angular integration is performed.
In particular we consider the case of the Galactic Center excess (GCE)
as given in
\cite{Calore:2014xka}\footnote{See also \cite{Daylan:2014rsa, Abazajian:2014fta, Gordon:2013vta,TheFermi-LAT:2015kwa}, and 
\cite{Petrovic:2014xra,Cholis:2015dea,Bartels:2015aea,Lee:2015fea} for an astrophysical interpretation of the excess.} where the authors consider a square of 40$^\circ \times$40$^\circ$ around
the GC, with a stripe of $\pm 2^\circ$ along the Galactic Plane excluded, for a total area
of 0.43 sr.
Fig.~\ref{fig:jfactor} shows the $\chi^2$ profile of the GCE $\mathcal{J}$-factor from our analysis for different cases
described in the caption. 
On technical note, we mention that when a derived parameter as $\mathcal{J}$ is involved,
our frequentist profiling methodology is slightly more involved.
In practice, first, for each point in our 7-d grid we derive the corresponding  $\mathcal{J}$ value.
Then, to build, for example, the 1-d profile $\chi^2$ for $\mathcal{J}$ we bin all the derived
$\mathcal{J}$ values in a new $\mathcal{J}$ grid. For each $\mathcal{J}$ bin we then take the minimum
$\chi^2$ among the $\chi^2$ corresponding to the $\mathcal{J}$ values  falling in that bin.
The final $\chi^2$ profile is shown in Fig.~\ref{fig:jfactor}.
This procedure can be easily generalized to 2-d cases, and is the more accurate the denser the original 7-d grid from which we start.   
Incidentally, we can see from Fig.~\ref{fig:jfactor} that the profile tend to a flat plateau
at low $\mathcal{J}$ with a $\Delta \chi^2$ with respect to the minimum of $\Delta \chi^2\sim 90-7=83$.
This, in practice, corresponds to the overall significance of our analysis
to the presence of DM in the Galaxy which is thus $\sqrt{\Delta \chi^2}\sim 9\, \sigma$.

The methodology can be easily extended to the calculation of $\mathcal{J}$-factors
over other regions for different analyses like GC searches for gamma-ray lines ~\cite{Abdalla:2016olq,Abdallah:2018qtu,Ackermann:2013uma,Ackermann:2015lka}
or DM searches at TeV with Cherenkov telescopes~\cite{Abdallah:2016ygi,HESS:2015cda,Acharya:2017ttl, Doro:2012xx} where the considered region
is of only few degrees, and thus even more sensitive to the uncertainties in the
DM distribution.

\subsection{Galactic Center Excess}

Given the $\mathcal{J}$-factor and its uncertainty (or $\chi^2$ profile), it is easy to include it in the GCE analysis.
The gamma-ray flux from the GCE is given by
\begin{equation}
    \frac{d\Phi}{dE} =    \frac{\langle \sigma v \rangle }{8\pi \, {m_\chi}^2} \frac{dN}{dE} \  \mathcal{J}   
    \label{eq:GCEflux}
\end{equation}
where $\langle \sigma v \rangle$ is the thermally averaged DM annihilation cross section,
$m_\chi$ the DM particle mass and $dN/dE$ is the spectrum of gamma-ray photons
from a single DM annihilation. 
For this last quantity we will use as example the case of annihilation into $b\bar{b}$ quarks
taking the spectrum from \cite{Cirelli:2010xx}.
We calculate the $\chi^2$ relative to the GCE as 
\begin{equation}
\chi^2_{\rm GCE} = \sum_{i, j}  (d_i - t_i) \left(\Sigma_{ij}+ \delta_{ij} (\sigma_{\rm rel}\, t_i)^2 \right)^{-1} (d_j - t_j),
\label{eq:chi2GCE}
\end{equation}
where $d_i$ are the GCE fluxes in the 24 energy bins given in \cite{Calore:2014xka},
$t_i$ is the model prediction from Eq.~\ref{eq:GCEflux}  (to be more precise, in \cite{Calore:2014xka}
fluxes are normalized to the area of the region analyzed, so from Eq.~\ref{eq:GCEflux} we further need
to divide by 0.43 sr),  and $\Sigma_{ij}$ is the covariance matrix among the energy bins,
again given in \cite{Calore:2014xka}.  
Similarly to what explained in \cite{Cuoco:2016jqt}, we further add to the covariance matrix a diagonal
error equal to $\sigma_{\rm rel}$ per cent of the model prediction $t_i$ to account
for the model uncertainty in the annihilation spectrum $dN/dE$. In particular,
as explained in \cite{Cuoco:2016jqt}, a choice of $\sigma_{\rm rel}=10\%$ is appropriate.

\begin{figure}[t]
\centering
\includegraphics[width=0.75\columnwidth]{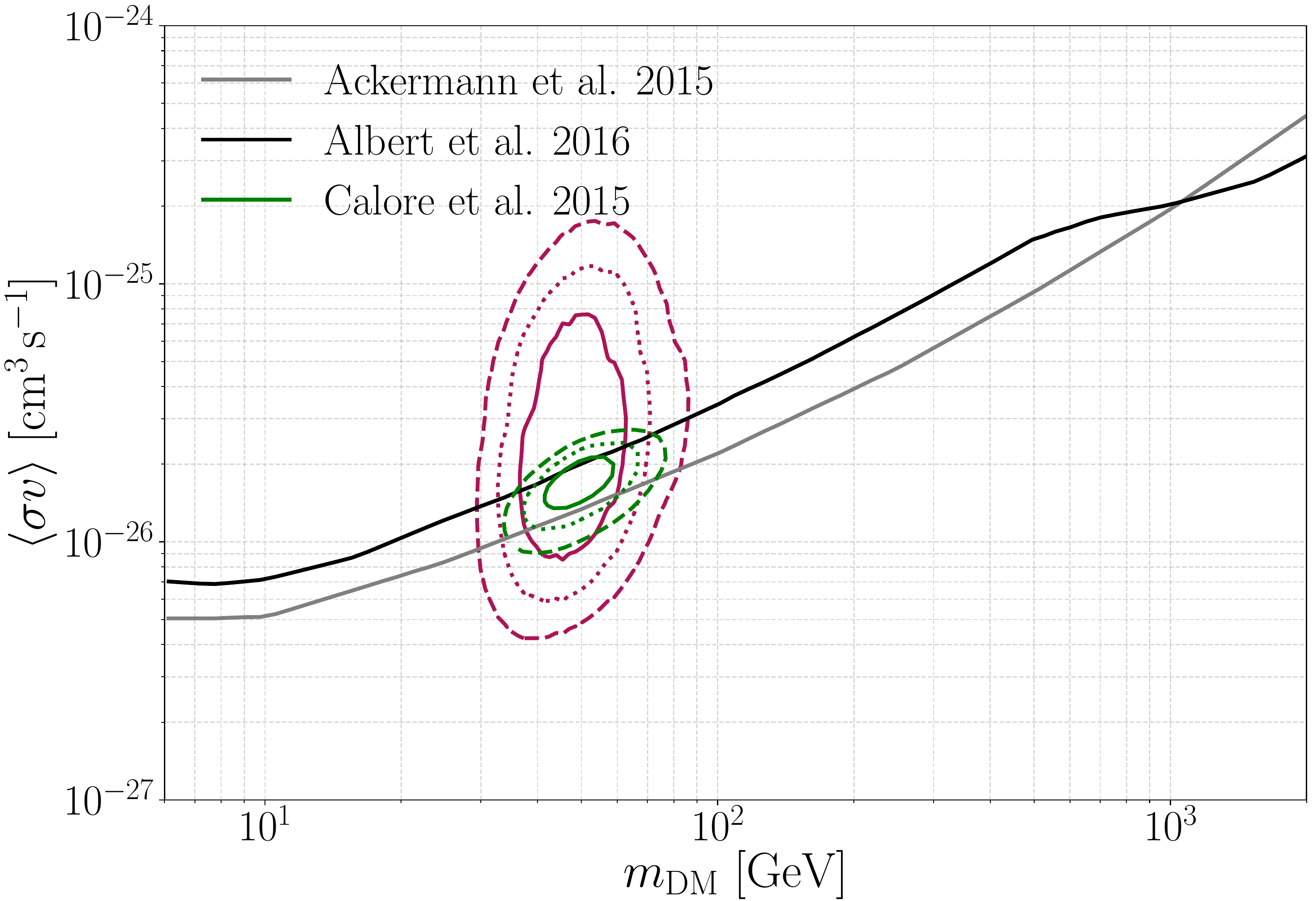}
\caption{1, 2 and 3 $\sigma$ contours in the DM particle parameter space $(m_{\rm DM}, \langle\sigma v\rangle)$ for the $b\bar{b}$ 
DM annihilation channel from our analysis, and
from the \cite{Calore:2014xka} work.    Also included are the upper limits from the analysis of
Milky Way dwarf galaxies \cite{Ackermann:2015zua,Fermi-LAT:2016uux}.
}
\label{fig:contour_mDMsigmav}
\end{figure}

To include in the GCE analysis the $\mathcal{J}$-factor uncertainties, we
consider a $\chi^2$ with 3 contributions:
\begin{equation}
\label{eq:chi2totshort}
\chi^2_{\rm total} = \chi^2_{\rm GCE} (\langle\sigma v\rangle, m_{\rm DM}, \mathcal{J}) + \chi^2_{\rm RC}(\mathcal{J}, \gamma) + \chi^2_{\gamma},
\end{equation}
where $\chi^2_{\rm GCE}$ is given by Eq.~\ref{eq:chi2GCE},
$\chi^2_{\rm RC}(\mathcal{J}, \gamma)$  is build from Eq.~\ref{eq:chi2RC} profiling
over $R_0$, $R_s$, $\rho_s$, $\langle\tau\rangle$, $\Sigma_*$ and $\mathcal{M}_i$ (but not $\gamma$) and
$\chi^2_{\gamma} = (\gamma-1.2)^2/0.08^2$ is a Gaussian prior on $\gamma$, with mean 1.2
and $\sigma=0.08$ again coming from the analysis of the morphology of the GCE in  \cite{Calore:2014xka}.
Constraints in the plane  $(m_{\rm DM}, \langle\sigma v\rangle)$ derived from 
 the above total $\chi^2$, further profiled over  $\gamma$ and $\mathcal{J}$,   are shown in Fig.~\ref{fig:contour_mDMsigmav}
and compared with the results of \cite{Calore:2014xka}.
We also cross-checked, for consistency, that fixing the $\mathcal{J}$-factor to the value adopted in \cite{Calore:2014xka}
we obtain their same $(m_{\rm DM}, \langle\sigma v\rangle)$ contours, which, for comparison are also shown in the same plot.
It can be seen that including the DM distribution uncertainties significantly enlarges the 
contours, both at large and small $ \langle\sigma v\rangle$.
In particular, while the preferred region of \cite{Calore:2014xka} is in (mild) tension with the null observations
of a gamma-ray signal from local dwarf galaxies~\cite{Ackermann:2015zua,Fermi-LAT:2016uux},
this tension disappear when considering the $\mathcal{J}$-factor uncertainties.
It should be mentioned, nonetheless, that the tension with dwarfs constraints
can also be further relieved if more conservative estimates of the DM
content of the dwarfs is adopted~\cite{Bonnivard:2015xpq}, or, similarly, if a more conservative 
analysis of the gamma-ray background at the dwarfs positions is performed~\cite{Calore:2018sdx,Mazziotta:2012ux}.

Simplified attempts to take into account the DM distribution uncertainties
for the GCE excess have been performed in \cite{Benito:2016kyp,Cuoco:2017rxb,Abazajian:2015raa,Keeley:2017fbz}.
In \cite{Benito:2016kyp}, similarly to here, variations with respect to the Galactic parameters are studied, but 
without performing a formal marginalization.
In \cite{Cuoco:2017rxb} the correlation among the Galactic parameters (in particular $\gamma$, $R_s$ and $\rho_s$) are extrapolated
from \cite{Nesti:2013uwa}, and the resulting GCE $\mathcal{J}$-factor is slightly overestimated with respect
to our results (contours in Fig.~\ref{fig:contour_mDMsigmav} reach  $ \langle\sigma v\rangle$ values a factor of two lower).
Finally, \cite{Abazajian:2015raa,Keeley:2017fbz}  uses correlations among the Galactic parameters
inferred from N-body simulations of MW-like DM haloes and priors on $\rho_0$ from
local analyses.

Finally, we mention that further priors on $R_0$ or $R_s, \rho_s$ or derived quantities like $\rho_0$
can be easily taken into account in our framework, if desired.
In this case, one needs to use the more general form of the $\chi^2$
\begin{equation}
\label{eq:chi2totlong}
\chi^2_{\rm total} = \chi^2_{\rm GCE} (\langle\sigma v\rangle, m_{\rm DM}, \mathcal{J}) + \chi^2_{\rm RC}(R_0, R_s, \rho_s, \gamma) + \chi^2_{\gamma} + 
\chi^2_{R_0, R_s, \rho_s},
\end{equation}
where  $ \chi^2_{\rm RC}$ is now profiled only over $\langle\tau\rangle$, $\Sigma_*$ and $\mathcal{M}_i$ and $\chi^2_{R_0, R_s, \rho_s}$
contains the other priors to be implemented. The $\mathcal{J}$-factor, this time is to be intended more
generally as a function of the DM parameters, $ \mathcal{J}=  \mathcal{J}(R_0, R_s, \rho_s, \gamma)$.
We derived the contours in the  $(m_{\rm DM}, \langle\sigma v\rangle)$ using this more general procedure
and using, as in the previous case, only the Gaussian prior on $\gamma$ and, as expected,
we obtained exactly the same results as using the simpler  $\chi^2$ version in Eq.~\ref{eq:chi2totshort}.

We provide the full $\chi^2_{\rm RC}(R_0, R_s, \rho_s, \gamma)$ table on a grid of the four
parameters. This represent the full information needed to reproduce the results of this section
and to specialize to specific cases of need. 
For completeness, we also provide a table containing  $\chi^2_{\rm RC}(\mathcal{J}, \gamma)$ 
although this can be derived from the first table using the procedure described above.
This second table can only be used for GCE analyses, though, since the $\mathcal{J}$-factor
refers to the GCE ROI, while the first table is completely general and can be used in any
analysis involving uncertainties in the Galactic DM distribution.

\subsection{Direct Detection}

The dependence of the results of direct searches for DM on the uncertainties of the properties
of the Galactic DM Halo are an active topic of research, 
see e.g.~\cite{Bhattacharjee2013, Bozorgnia2017, Bozorgnia2016, Fairbairn2013, Fornasa2014, Frandsen2012, Green2017, Green2012, Kuhlen2010, Peter2010, Pillepich2014, Vogelsberger2009}.
Results of direct searches depends on the DM Halo in two ways.
The main dependence is typically from $\rho_0$, the local DM density,
which enters linearly in the expected DM detection rate (see e.g.~\cite{Cerdeno:2010jj}).
Thus, the typical exclusion limits of the  DM-proton scattering cross-section  as function of DM mass
(see, e.g., the recent XENON1T experiment results \cite{2018arXiv180512562A}), can
just be linearly rescaled to a new $\rho_0$ value and weighted according to the $\rho_0$ likelihood.
The second dependence is from the velocity distribution of DM particles $f(v)$.
Assuming this distribution is  a Maxwelliann one can simply express the 
velocity dispersion parameter, $\sigma_v$, entering in the Maxwellian 
as $\sigma_v = \sqrt{3/2} V_0$. Thus, under the  Maxwellian  approximation,
our results can be used to take into account the uncertainties in $f(v)$
through the uncertainty in $V_0$.
We do not attempt here to include the effect of the variation of $V_0$, although this
can be implemented starting from our provided $\chi^2_{\rm RC}$ table.

Dropping the assumption of Maxwellian distribution can have major effects
on the direct DM constraints, the larger the deviation of $f(v)$ from a
Maxwellian  (see the recent study \cite{Ibarra:2018yxq}).
However, under the reasonable assumption of isotropic DM velocity and of system
at equilibrium $f(v)$ cannot be arbitrary but has to satisfy constraints given 
by the DM spatial distribution and the Boltzmann equation.
This strategy to constrain $f(v)$ has been indeed pursued
in various studies   \cite{Bhattacharjee2013, Fairbairn2013, Fornasa2014}.
Again, these kind of studies can  be in principle gereralized including the DM distribution uncertainties
tabulated in the present study to derive the related uncertainties on the reconstructed $f(v)$,             
i.e., in practice, propagating  the DM distribution uncertainties into the velocity distribution.


\subsection{Combined Fits}

\begin{figure}[t]
\centering
\includegraphics[width=0.75\columnwidth]{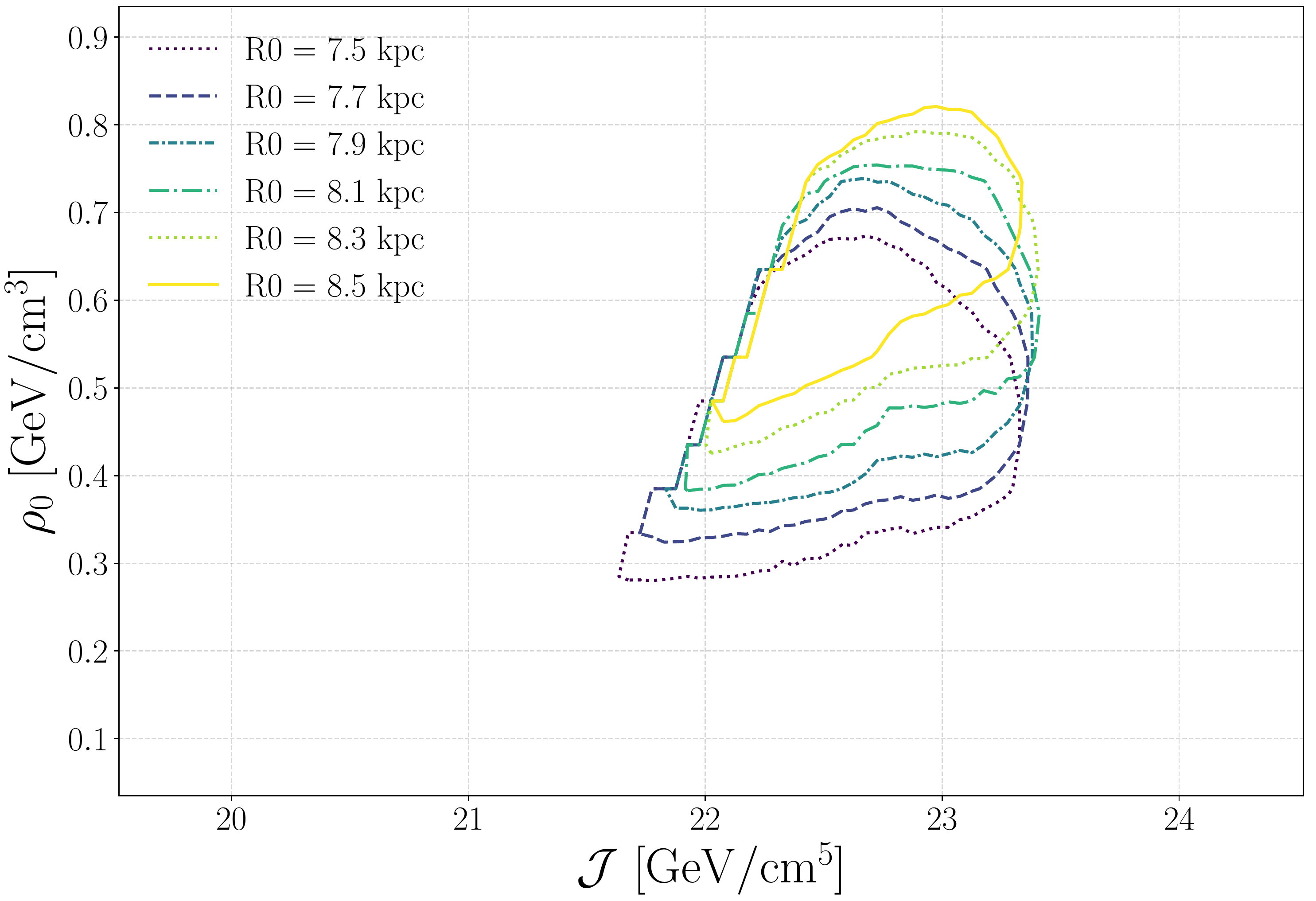}
\caption{Degeneracy between the GCE $\mathcal{J}$-factor and the local DM density
$\rho_0$ broken down into the contribution for different values of $R_0$.}
\label{fig:contourJrho0}
\end{figure}

Finally, a more subtle effect can appear when performing combined fits of GC constraints
or hints of signals like the GCE and local observations like the above direct detection constraints
or, for example, antiproton constraints (see e.g., \cite{Cuoco:2017rxb}).
In this case correlations might appear between the two observables.
This is illustrated in Fig.~\ref{fig:contourJrho0} which shows that
indeed there is a degeneracy between the GCE $\mathcal{J}$-factor and
$\rho_0$, mainly produced by the variation of $R_0$.
Again this can be taken into account using our tabulated likelihood.

\section{Summary and Conclusion}
We have used the observed Rotation Curve of the Milky Way up to $\approx 25$ kpc in Galactocentric radius
to constrain the parameters of a generalized NFW Dark Matter density profile.
We have improved with respect to previous analyses in several ways.
First, we have adopted a systematic statistical approach, scanning
the relevant parameter space and accurately exploring the various
degeneracies present. This last point is particularly important 
since several degeneracies exist, and precisely mapping them
is necessary to have reliable final error estimates.
Second, we use an accurate treatment of the
systematic uncertainties arising from the modeling of the visible components of the MW,
by both considering different baryonic morphologies for the Disc and the Bulge,
and allowing for each morphology mass variations within the uncertainties
given by microlensing and stellar surface-density measurements.
These baryonic uncertainties are fully marginalized (profiled) away
within our statistical framework.
We find that  the local DM density $\rho_0$ is constrained to the range $0.3-0.8\;{\rm GeV/cm^3}$ at the $2\sigma$ level, 
showing a strong positive correlation with the Sun's Galactocentric distance $R_0$.
The inner slope of the DM profile, $\gamma$, is very weakly constrained and 
both core ($\gamma\sim 0$) and cusp ($\gamma\sim 1$) DM density profiles are allowed.
Some combination of parameters can be, however, strongly constrained. For example
the often used standard $\rho_0=0.3$ GeV/cm$^3$, $R_0=8.5$ kpc is
disfavored at more than 4 $\sigma$. 
We release the likelihood of our analysis, namely a 4-dimensional table
listing $\chi^2$ values over a grid in $\gamma$, $R_0$, $R_s$, $\rho_s$, 
the latter two parameters being the scale radius and density scale of
the generalized NFW profile. In the above likelihood the baryonic physics
and related uncertainties have been already profiled away.
We have provided some example for the use of the likelihood,
in particular we have employed it in the analysis of the Galactic Center gamma-ray 
excess. We have found that the uncertainties in the DM profile 
significantly enlarge the allowed cross-section range, by a factor 3 to 4.
Other contexts in which our tabulated likelihood can be employed
involve Galactic Center or Galactic Halo DM searches in gamma rays
at GeV energies or TeV with Cherenkov telescopes,  DM neutrinos searches,
direct DM searches, local DM searches with antimatter, and combined local and GC
searches.

\section*{Acknowledgements}
F.~I.~acknowledges support from the Simons Foundation and FAPESP process 2014/11070-2.
This research was supported by resources supplied by the Center for Scientific Computing (NCC/GridUNESP) of the S\~ao Paulo State University (UNESP).
This work has been possible following the visit of A.C. to ICTP-SAIFR in S\~ao Paulo, supported by a Theodore von K\'arm\'an  fellowship
from RWTH University of Aachen, Germany.
We wish to thank Gabrijela Zaharias, Francesca Calore, Roberto Trotta and Jan Heisig for useful comments on the manuscript.

\bibliographystyle{JHEP}
\bibliography{freq-like-MWunc}

\providecommand{\href}[2]{#2}\begingroup\raggedright\begin{thebibliography}{10}

\bibitem{Navarro:1996gj}
J.~F. Navarro, C.~S. Frenk and S.~D.~M. White, \emph{{A Universal density
  profile from hierarchical clustering}},
  \href{http://dx.doi.org/10.1086/304888}{\emph{Astrophys. J.} {\bf 490} (1997)
  493--508}, [\href{https://arxiv.org/abs/astro-ph/9611107}{{\tt
  astro-ph/9611107}}].

\bibitem{Burkert:1995yz}
A.~Burkert, \emph{{The Structure of dark matter halos in dwarf galaxies}},
  \href{http://dx.doi.org/10.1086/309560}{\emph{IAU Symp.} {\bf 171} (1996)
  175}, [\href{https://arxiv.org/abs/astro-ph/9504041}{{\tt
  astro-ph/9504041}}].

\bibitem{Einasto1965}
J.~{Einasto}, \emph{{On the Construction of a Composite Model for the Galaxy
  and on the Determination of the System of Galactic Parameters}}, {\emph{Trudy
  Astrofizicheskogo Instituta Alma-Ata} {\bf 5} (1965) 87--100}.

\bibitem{Caldwell:1981}
J.~A.~R. {Caldwell} and J.~P. {Ostriker}, \emph{{The mass distribution within
  our Galaxy - A three component model}},
  \href{http://dx.doi.org/10.1086/159441}{\emph{\apj} {\bf 251} (Dec., 1981)
  61--87}.

\bibitem{Iocco2011}
F.~{Iocco}, M.~{Pato}, G.~{Bertone} and P.~{Jetzer}, \emph{{Dark Matter
  distribution in the Milky Way: microlensing and dynamical constraints}},
  \href{http://dx.doi.org/10.1088/1475-7516/2011/11/029}{\emph{\jcap} {\bf 11}
  (Nov., 2011) 029}, [\href{https://arxiv.org/abs/1107.5810}{{\tt 1107.5810}}].

\bibitem{Iocco2015}
F.~{Iocco}, M.~{Pato} and G.~{Bertone}, \emph{{Evidence for dark matter in the
  inner Milky Way}}, \href{http://dx.doi.org/10.1038/nphys3237}{\emph{Nature
  Physics} {\bf 11} (Mar., 2015) 245--248},
  [\href{https://arxiv.org/abs/1502.03821}{{\tt 1502.03821}}].

\bibitem{Pato:2015dua}
M.~Pato, F.~Iocco and G.~Bertone, \emph{{Dynamical constraints on the dark
  matter distribution in the Milky Way}},
  \href{http://dx.doi.org/10.1088/1475-7516/2015/12/001}{\emph{JCAP} {\bf 1512}
  (2015) 001}, [\href{https://arxiv.org/abs/1504.06324}{{\tt 1504.06324}}].

\bibitem{Sofue:2008wt}
Y.~Sofue, M.~Honma and T.~Omodaka, \emph{{Unified Rotation Curve of the Galaxy
  -- Decomposition into de Vaucouleurs Bulge, Disk, Dark Halo, and the 9-kpc
  Rotation Dip --}}, \href{http://dx.doi.org/10.1093/pasj/61.2.227}{\emph{Publ.
  Astron. Soc. Jap.} {\bf 61} (2009) 227},
  [\href{https://arxiv.org/abs/0811.0859}{{\tt 0811.0859}}].

\bibitem{Sofue2012}
Y.~{Sofue}, \emph{{Grand Rotation Curve and Dark Matter Halo in the Milky Way
  Galaxy}}, \href{http://dx.doi.org/10.1093/pasj/64.4.75}{\emph{\pasj} {\bf 64}
  (Aug., 2012) 75}, [\href{https://arxiv.org/abs/1110.4431}{{\tt 1110.4431}}].

\bibitem{CatenaUllio2010}
R.~{Catena} and P.~{Ullio}, \emph{{A novel determination of the local dark
  matter density}},
  \href{http://dx.doi.org/10.1088/1475-7516/2010/08/004}{\emph{\jcap} {\bf 8}
  (Aug., 2010) 004}, [\href{https://arxiv.org/abs/0907.0018}{{\tt 0907.0018}}].

\bibitem{Nesti:2013uwa}
F.~Nesti and P.~Salucci, \emph{{The Dark Matter halo of the Milky Way, AD
  2013}}, \href{http://dx.doi.org/10.1088/1475-7516/2013/07/016}{\emph{JCAP}
  {\bf 1307} (2013) 016}, [\href{https://arxiv.org/abs/1304.5127}{{\tt
  1304.5127}}].

\bibitem{Huang2016}
Y.~{Huang}, X.-W. {Liu}, H.-B. {Yuan}, M.-S. {Xiang}, H.-W. {Zhang}, B.-Q.
  {Chen} et~al., \emph{{The Milky Way's rotation curve out to 100 kpc and its
  constraint on the Galactic mass distribution}},
  \href{http://dx.doi.org/10.1093/mnras/stw2096}{\emph{\mnras} {\bf 463} (Dec.,
  2016) 2623--2639}, [\href{https://arxiv.org/abs/1604.01216}{{\tt
  1604.01216}}].

\bibitem{McMillan2011}
P.~J. {McMillan}, \emph{{Mass models of the Milky Way}},
  \href{http://dx.doi.org/10.1111/j.1365-2966.2011.18564.x}{\emph{\mnras} {\bf
  414} (July, 2011) 2446--2457}, [\href{https://arxiv.org/abs/1102.4340}{{\tt
  1102.4340}}].

\bibitem{McMillan2017}
P.~J. {McMillan}, \emph{{The mass distribution and gravitational potential of
  the Milky Way}}, \href{http://dx.doi.org/10.1093/mnras/stw2759}{\emph{\mnras}
  {\bf 465} (Feb., 2017) 76--94}, [\href{https://arxiv.org/abs/1608.00971}{{\tt
  1608.00971}}].

\bibitem{DeBoer2009}
M.~{Weber} and W.~{de Boer}, \emph{{Determination of the local dark matter
  density in our Galaxy}},
  \href{http://dx.doi.org/10.1051/0004-6361/200913381}{\emph{\aap} {\bf 509}
  (Jan., 2010) A25}, [\href{https://arxiv.org/abs/0910.4272}{{\tt 0910.4272}}].

\bibitem{Xue2008}
X.~X. {Xue}, H.~W. {Rix}, G.~{Zhao}, P.~{Re Fiorentin}, T.~{Naab},
  M.~{Steinmetz} et~al., \emph{{The Milky Way's Circular Velocity Curve to 60
  kpc and an Estimate of the Dark Matter Halo Mass from the Kinematics of
  \~{}2400 SDSS Blue Horizontal-Branch Stars}},
  \href{http://dx.doi.org/10.1086/589500}{\emph{\apj} {\bf 684} (Sept., 2008)
  1143--1158}, [\href{https://arxiv.org/abs/0801.1232}{{\tt 0801.1232}}].

\bibitem{Bovy2009}
J.~{Bovy}, D.~W. {Hogg} and H.-W. {Rix}, \emph{{Galactic Masers and the Milky
  Way Circular Velocity}},
  \href{http://dx.doi.org/10.1088/0004-637X/704/2/1704}{\emph{\apj} {\bf 704}
  (Oct., 2009) 1704--1709}, [\href{https://arxiv.org/abs/0907.5423}{{\tt
  0907.5423}}].

\bibitem{BovyAPOGEE2012}
J.~{Bovy}, C.~{Allende Prieto}, T.~C. {Beers}, D.~{Bizyaev}, L.~N. {da Costa},
  K.~{Cunha} et~al., \emph{{The Milky Way's Circular-velocity Curve between 4
  and 14 kpc from APOGEE data}},
  \href{http://dx.doi.org/10.1088/0004-637X/759/2/131}{\emph{\apj} {\bf 759}
  (Nov., 2012) 131}, [\href{https://arxiv.org/abs/1209.0759}{{\tt 1209.0759}}].

\bibitem{Abdalla:2016olq}
{\scshape H.E.S.S.} collaboration, H.~Abdalla et~al., \emph{{H.E.S.S. Limits on
  Linelike Dark Matter Signatures in the 100 GeV to 2 TeV Energy Range Close to
  the Galactic Center}},
  \href{http://dx.doi.org/10.1103/PhysRevLett.117.151302}{\emph{Phys. Rev.
  Lett.} {\bf 117} (2016) 151302},
  [\href{https://arxiv.org/abs/1609.08091}{{\tt 1609.08091}}].

\bibitem{Abdallah:2018qtu}
{\scshape HESS} collaboration, H.~Abdallah et~al., \emph{{Search for
  $\gamma$-Ray Line Signals from Dark Matter Annihilations in the Inner
  Galactic Halo from 10 Years of Observations with H.E.S.S.}},
  \href{http://dx.doi.org/10.1103/PhysRevLett.120.201101}{\emph{Phys. Rev.
  Lett.} {\bf 120} (2018) 201101},
  [\href{https://arxiv.org/abs/1805.05741}{{\tt 1805.05741}}].

\bibitem{Ackermann:2013uma}
{\scshape Fermi-LAT} collaboration, M.~Ackermann et~al., \emph{{Search for
  Gamma-ray Spectral Lines with the Fermi Large Area Telescope and Dark Matter
  Implications}},
  \href{http://dx.doi.org/10.1103/PhysRevD.88.082002}{\emph{Phys. Rev.} {\bf
  D88} (2013) 082002}, [\href{https://arxiv.org/abs/1305.5597}{{\tt
  1305.5597}}].

\bibitem{Ackermann:2015lka}
{\scshape Fermi-LAT} collaboration, M.~Ackermann et~al., \emph{{Updated search
  for spectral lines from Galactic dark matter interactions with pass 8 data
  from the Fermi Large Area Telescope}},
  \href{http://dx.doi.org/10.1103/PhysRevD.91.122002}{\emph{Phys. Rev.} {\bf
  D91} (2015) 122002}, [\href{https://arxiv.org/abs/1506.00013}{{\tt
  1506.00013}}].

\bibitem{Calore:2014xka}
F.~Calore, I.~Cholis and C.~Weniger, \emph{{Background Model Systematics for
  the Fermi GeV Excess}},
  \href{http://dx.doi.org/10.1088/1475-7516/2015/03/038}{\emph{JCAP} {\bf 1503}
  (2015) 038}, [\href{https://arxiv.org/abs/1409.0042}{{\tt 1409.0042}}].

\bibitem{TheFermi-LAT:2017vmf}
{\scshape Fermi-LAT} collaboration, M.~Ackermann et~al., \emph{{The Fermi
  Galactic Center GeV Excess and Implications for Dark Matter}},
  \href{http://dx.doi.org/10.3847/1538-4357/aa6cab}{\emph{Astrophys. J.} {\bf
  840} (2017) 43}, [\href{https://arxiv.org/abs/1704.03910}{{\tt 1704.03910}}].

\bibitem{Abdallah:2016ygi}
{\scshape H.E.S.S.} collaboration, H.~Abdallah et~al., \emph{{Search for dark
  matter annihilations towards the inner Galactic halo from 10 years of
  observations with H.E.S.S}},
  \href{http://dx.doi.org/10.1103/PhysRevLett.117.111301}{\emph{Phys. Rev.
  Lett.} {\bf 117} (2016) 111301},
  [\href{https://arxiv.org/abs/1607.08142}{{\tt 1607.08142}}].

\bibitem{HESS:2015cda}
{\scshape H.E.S.S.} collaboration, A.~Abramowski et~al., \emph{{Constraints on
  an Annihilation Signal from a Core of Constant Dark Matter Density around the
  Milky Way Center with H.E.S.S.}},
  \href{http://dx.doi.org/10.1103/PhysRevLett.114.081301}{\emph{Phys. Rev.
  Lett.} {\bf 114} (2015) 081301},
  [\href{https://arxiv.org/abs/1502.03244}{{\tt 1502.03244}}].

\bibitem{Acharya:2017ttl}
{\scshape Cherenkov Telescope Array Consortium} collaboration, B.~S. Acharya
  et~al., \emph{{Science with the Cherenkov Telescope Array}},
  \href{https://arxiv.org/abs/1709.07997}{{\tt 1709.07997}}.

\bibitem{Doro:2012xx}
{\scshape CTA Consortium} collaboration, M.~Doro et~al., \emph{{Dark Matter and
  Fundamental Physics with the Cherenkov Telescope Array}},
  \href{http://dx.doi.org/10.1016/j.astropartphys.2012.08.002}{\emph{Astropart.
  Phys.} {\bf 43} (2013) 189--214},
  [\href{https://arxiv.org/abs/1208.5356}{{\tt 1208.5356}}].

\bibitem{Aartsen:2018mxl}
{\scshape IceCube} collaboration, M.~G. Aartsen et~al., \emph{{Search for
  neutrinos from decaying dark matter with IceCube}},
  \href{https://arxiv.org/abs/1804.03848}{{\tt 1804.03848}}.

\bibitem{Aartsen:2017ulx}
{\scshape IceCube} collaboration, M.~G. Aartsen et~al., \emph{{Search for
  Neutrinos from Dark Matter Self-Annihilations in the center of the Milky Way
  with 3 years of IceCube/DeepCore}},
  \href{http://dx.doi.org/10.1140/epjc/s10052-017-5213-y}{\emph{Eur. Phys. J.}
  {\bf C77} (2017) 627}, [\href{https://arxiv.org/abs/1705.08103}{{\tt
  1705.08103}}].

\bibitem{Aartsen:2015xej}
{\scshape IceCube} collaboration, M.~G. Aartsen et~al., \emph{{Search for Dark
  Matter Annihilation in the Galactic Center with IceCube-79}},
  \href{http://dx.doi.org/10.1140/epjc/s10052-015-3713-1}{\emph{Eur. Phys. J.}
  {\bf C75} (2015) 492}, [\href{https://arxiv.org/abs/1505.07259}{{\tt
  1505.07259}}].

\bibitem{Adrian-Martinez:2015wey}
{\scshape ANTARES} collaboration, S.~Adrian-Martinez et~al., \emph{{Search of
  Dark Matter Annihilation in the Galactic Centre using the ANTARES Neutrino
  Telescope}},
  \href{http://dx.doi.org/10.1088/1475-7516/2015/10/068}{\emph{JCAP} {\bf 1510}
  (2015) 068}, [\href{https://arxiv.org/abs/1505.04866}{{\tt 1505.04866}}].

\bibitem{Cuoco:2016eej}
A.~Cuoco, M.~KrŠmer and M.~Korsmeier, \emph{{Novel Dark Matter Constraints from
  Antiprotons in Light of AMS-02}},
  \href{http://dx.doi.org/10.1103/PhysRevLett.118.191102}{\emph{Phys. Rev.
  Lett.} {\bf 118} (2017) 191102},
  [\href{https://arxiv.org/abs/1610.03071}{{\tt 1610.03071}}].

\bibitem{Giesen:2015ufa}
G.~Giesen, M.~Boudaud, Y.~GŽnolini, V.~Poulin, M.~Cirelli, P.~Salati et~al.,
  \emph{{AMS-02 antiprotons, at last! Secondary astrophysical component and
  immediate implications for Dark Matter}},
  \href{http://dx.doi.org/10.1088/1475-7516/2015/09/023,
  10.1088/1475-7516/2015/9/023}{\emph{JCAP} {\bf 1509} (2015) 023},
  [\href{https://arxiv.org/abs/1504.04276}{{\tt 1504.04276}}].

\bibitem{Fornengo:2013xda}
N.~Fornengo, L.~Maccione and A.~Vittino, \emph{{Constraints on particle dark
  matter from cosmic-ray antiprotons}},
  \href{http://dx.doi.org/10.1088/1475-7516/2014/04/003}{\emph{JCAP} {\bf 1404}
  (2014) 003}, [\href{https://arxiv.org/abs/1312.3579}{{\tt 1312.3579}}].

\bibitem{Donato:2003xg}
F.~Donato, N.~Fornengo, D.~Maurin and P.~Salati, \emph{{Antiprotons in cosmic
  rays from neutralino annihilation}},
  \href{http://dx.doi.org/10.1103/PhysRevD.69.063501}{\emph{Phys. Rev.} {\bf
  D69} (2004) 063501}, [\href{https://arxiv.org/abs/astro-ph/0306207}{{\tt
  astro-ph/0306207}}].

\bibitem{Hooper:2014ysa}
D.~Hooper, T.~Linden and P.~Mertsch, \emph{{What Does The PAMELA Antiproton
  Spectrum Tell Us About Dark Matter?}},
  \href{http://dx.doi.org/10.1088/1475-7516/2015/03/021}{\emph{JCAP} {\bf 1503}
  (2015) 021}, [\href{https://arxiv.org/abs/1410.1527}{{\tt 1410.1527}}].

\bibitem{Bringmann:2006im}
T.~Bringmann and P.~Salati, \emph{{The galactic antiproton spectrum at high
  energies: Background expectation vs. exotic contributions}},
  \href{http://dx.doi.org/10.1103/PhysRevD.75.083006}{\emph{Phys. Rev.} {\bf
  D75} (2007) 083006}, [\href{https://arxiv.org/abs/astro-ph/0612514}{{\tt
  astro-ph/0612514}}].

\bibitem{Gaggero:2018zbd}
D.~Gaggero and M.~Valli, \emph{{Impact of cosmic-ray physics on dark matter
  indirect searches}},  \href{https://arxiv.org/abs/1802.00636}{{\tt
  1802.00636}}.

\bibitem{Bhattacharjee2013}
P.~{Bhattacharjee}, S.~{Chaudhury}, S.~{Kundu} and S.~{Majumdar},
  \emph{{Deriving the velocity distribution of Galactic dark matter particles
  from the rotation curve data}},
  \href{http://dx.doi.org/10.1103/PhysRevD.87.083525}{\emph{\prd} {\bf 87}
  (Apr., 2013) 083525}, [\href{https://arxiv.org/abs/1210.2328}{{\tt
  1210.2328}}].

\bibitem{Bozorgnia2017}
N.~{Bozorgnia} and G.~{Bertone}, \emph{{Implications of hydrodynamical
  simulations for the interpretation of direct dark matter searches}},
  \href{http://dx.doi.org/10.1142/S0217751X17300162}{\emph{International
  Journal of Modern Physics A} {\bf 32} (July, 2017) 1730016},
  [\href{https://arxiv.org/abs/1705.05853}{{\tt 1705.05853}}].

\bibitem{Bozorgnia2016}
N.~{Bozorgnia}, F.~{Calore}, M.~{Schaller}, M.~{Lovell}, G.~{Bertone}, C.~S.
  {Frenk} et~al., \emph{{Simulated Milky Way analogues: implications for dark
  matter direct searches}},
  \href{http://dx.doi.org/10.1088/1475-7516/2016/05/024}{\emph{\jcap} {\bf 5}
  (May, 2016) 024}, [\href{https://arxiv.org/abs/1601.04707}{{\tt
  1601.04707}}].

\bibitem{Fairbairn2013}
M.~{Fairbairn}, T.~{Douce} and J.~{Swift}, \emph{{Quantifying astrophysical
  uncertainties on dark matter direct detection results}},
  \href{http://dx.doi.org/10.1016/j.astropartphys.2013.06.003}{\emph{Astroparticle
  Physics} {\bf 47} (July, 2013) 45--53},
  [\href{https://arxiv.org/abs/1206.2693}{{\tt 1206.2693}}].

\bibitem{Fornasa2014}
M.~{Fornasa} and A.~M. {Green}, \emph{{Self-consistent phase-space distribution
  function for the anisotropic dark matter halo of the Milky{\^A} Way}},
  \href{http://dx.doi.org/10.1103/PhysRevD.89.063531}{\emph{\prd} {\bf 89}
  (Mar., 2014) 063531}, [\href{https://arxiv.org/abs/1311.5477}{{\tt
  1311.5477}}].

\bibitem{Frandsen2012}
M.~T. {Frandsen}, F.~{Kahlhoefer}, C.~{McCabe}, S.~{Sarkar} and
  K.~{Schmidt-Hoberg}, \emph{{Resolving astrophysical uncertainties in dark
  matter direct detection}},
  \href{http://dx.doi.org/10.1088/1475-7516/2012/01/024}{\emph{\jcap} {\bf 1}
  (Jan., 2012) 024}, [\href{https://arxiv.org/abs/1111.0292}{{\tt 1111.0292}}].

\bibitem{Green2017}
A.~M. {Green}, \emph{{Astrophysical uncertainties on the local dark matter
  distribution and direct detection experiments}},
  \href{http://dx.doi.org/10.1088/1361-6471/aa7819}{\emph{Journal of Physics G
  Nuclear Physics} {\bf 44} (Aug., 2017) 084001},
  [\href{https://arxiv.org/abs/1703.10102}{{\tt 1703.10102}}].

\bibitem{Green2012}
A.~M. {Green}, \emph{{Astrophysical Uncertainties on Direct Detection
  Experiments}},
  \href{http://dx.doi.org/10.1142/S0217732312300042}{\emph{Modern Physics
  Letters A} {\bf 27} (2012) 1230004--1--1230004--20},
  [\href{https://arxiv.org/abs/1112.0524}{{\tt 1112.0524}}].

\bibitem{Kuhlen2010}
M.~{Kuhlen}, N.~{Weiner}, J.~{Diemand}, P.~{Madau}, B.~{Moore}, D.~{Potter}
  et~al., \emph{{Dark matter direct detection with non-Maxwellian velocity
  structure}},
  \href{http://dx.doi.org/10.1088/1475-7516/2010/02/030}{\emph{\jcap} {\bf 2}
  (Feb., 2010) 030}, [\href{https://arxiv.org/abs/0912.2358}{{\tt 0912.2358}}].

\bibitem{Peter2010}
A.~H.~G. {Peter}, \emph{{Getting the astrophysics and particle physics of dark
  matter out of next-generation direct detection experiments}},
  \href{http://dx.doi.org/10.1103/PhysRevD.81.087301}{\emph{\prd} {\bf 81}
  (Apr., 2010) 087301}, [\href{https://arxiv.org/abs/0910.4765}{{\tt
  0910.4765}}].

\bibitem{Pillepich2014}
A.~{Pillepich}, M.~{Kuhlen}, J.~{Guedes} and P.~{Madau}, \emph{{The
  Distribution of Dark Matter in the Milky Way's Disk}},
  \href{http://dx.doi.org/10.1088/0004-637X/784/2/161}{\emph{\apj} {\bf 784}
  (Apr., 2014) 161}, [\href{https://arxiv.org/abs/1308.1703}{{\tt 1308.1703}}].

\bibitem{Vogelsberger2009}
M.~{Vogelsberger}, A.~{Helmi}, V.~{Springel}, S.~D.~M. {White}, J.~{Wang},
  C.~S. {Frenk} et~al., \emph{{Phase-space structure in the local dark matter
  distribution and its signature in direct detection experiments}},
  \href{http://dx.doi.org/10.1111/j.1365-2966.2009.14630.x}{\emph{\mnras} {\bf
  395} (May, 2009) 797--811}, [\href{https://arxiv.org/abs/0812.0362}{{\tt
  0812.0362}}].

\bibitem{Bienayme2014}
O.~{Bienaym{\'e}}, B.~{Famaey}, A.~{Siebert}, K.~C. {Freeman}, B.~K. {Gibson},
  G.~{Gilmore} et~al., \emph{{Weighing the local dark matter with RAVE red
  clump stars}},
  \href{http://dx.doi.org/10.1051/0004-6361/201424478}{\emph{\aap} {\bf 571}
  (Nov., 2014) A92}, [\href{https://arxiv.org/abs/1406.6896}{{\tt 1406.6896}}].

\bibitem{Bovy2012}
J.~{Bovy} and S.~{Tremaine}, \emph{{On the Local Dark Matter Density}},
  \href{http://dx.doi.org/10.1088/0004-637X/756/1/89}{\emph{\apj} {\bf 756}
  (Sept., 2012) 89}, [\href{https://arxiv.org/abs/1205.4033}{{\tt 1205.4033}}].

\bibitem{Garbari2012}
S.~{Garbari}, C.~{Liu}, J.~I. {Read} and G.~{Lake}, \emph{{A new determination
  of the local dark matter density from the kinematics of K dwarfs}},
  \href{http://dx.doi.org/10.1111/j.1365-2966.2012.21608.x}{\emph{\mnras} {\bf
  425} (Sept., 2012) 1445--1458}, [\href{https://arxiv.org/abs/1206.0015}{{\tt
  1206.0015}}].

\bibitem{McKee2015}
C.~F. {McKee}, A.~{Parravano} and D.~J. {Hollenbach}, \emph{{Stars, Gas, and
  Dark Matter in the Solar Neighborhood}},
  \href{http://dx.doi.org/10.1088/0004-637X/814/1/13}{\emph{\apj} {\bf 814}
  (Nov., 2015) 13}, [\href{https://arxiv.org/abs/1509.05334}{{\tt
  1509.05334}}].

\bibitem{Salucci2010}
P.~{Salucci}, F.~{Nesti}, G.~{Gentile} and C.~{Frigerio Martins}, \emph{{The
  dark matter density at the Sun's location}},
  \href{http://dx.doi.org/10.1051/0004-6361/201014385}{\emph{\aap} {\bf 523}
  (Nov., 2010) A83}, [\href{https://arxiv.org/abs/1003.3101}{{\tt 1003.3101}}].

\bibitem{Sivertsson2018}
S.~{Sivertsson}, H.~{Silverwood}, J.~I. {Read}, G.~{Bertone} and P.~{Steger},
  \emph{{The localdark matter density from SDSS-SEGUE G-dwarfs}},
  \href{http://dx.doi.org/10.1093/mnras/sty977}{\emph{\mnras} {\bf 478} (Aug.,
  2018) 1677--1693}, [\href{https://arxiv.org/abs/1708.07836}{{\tt
  1708.07836}}].

\bibitem{XiaLAMOST2016}
Q.~{Xia}, C.~{Liu}, S.~{Mao}, Y.~{Song}, L.~{Zhang}, R.~J. {Long} et~al.,
  \emph{{Determining the local dark matter density with LAMOST data}},
  \href{http://dx.doi.org/10.1093/mnras/stw565}{\emph{\mnras} {\bf 458} (June,
  2016) 3839--3850}, [\href{https://arxiv.org/abs/1510.06810}{{\tt
  1510.06810}}].

\bibitem{Zhang2013}
L.~{Zhang}, H.-W. {Rix}, G.~{van de Ven}, J.~{Bovy}, C.~{Liu} and G.~{Zhao},
  \emph{{The Gravitational Potential near the Sun from SEGUE K-dwarf
  Kinematics}},
  \href{http://dx.doi.org/10.1088/0004-637X/772/2/108}{\emph{\apj} {\bf 772}
  (Aug., 2013) 108}, [\href{https://arxiv.org/abs/1209.0256}{{\tt 1209.0256}}].

\bibitem{Read2014}
J.~I. {Read}, \emph{{The local dark matter density}},
  \href{http://dx.doi.org/10.1088/0954-3899/41/6/063101}{\emph{Journal of
  Physics G Nuclear Physics} {\bf 41} (June, 2014) 063101},
  [\href{https://arxiv.org/abs/1404.1938}{{\tt 1404.1938}}].

\bibitem{Pato:2017yai}
M.~Pato and F.~Iocco, \emph{{{\tt galkin}: a new compilation of the Milky Way
  rotation curve data}},
  \href{http://dx.doi.org/10.1016/j.softx.2016.12.006}{\emph{SoftwareX} {\bf
  Volume 6} (2017) }, [\href{https://arxiv.org/abs/1703.00020}{{\tt
  1703.00020}}].

\bibitem{Ferriere:2007yq}
K.~Ferriere, W.~Gillard and P.~Jean, \emph{{Spatial distribution of
  interstellar gas in the innermost 3 kpc of our Galaxy}},
  \href{http://dx.doi.org/10.1051/0004-6361:20066992}{\emph{Astron. Astrophys.}
  {\bf 467} (2007) 611--627},
  [\href{https://arxiv.org/abs/astro-ph/0702532}{{\tt astro-ph/0702532}}].

\bibitem{Ferriere:1998}
K.~{Ferri{\`e}re}, \emph{{Global Model of the Interstellar Medium in Our Galaxy
  with New Constraints on the Hot Gas Component}},
  \href{http://dx.doi.org/10.1086/305469}{\emph{\apj} {\bf 497} (Apr., 1998)
  759--776}.

\bibitem{Popowski:2004uv}
{\scshape MACHO} collaboration, P.~Popowski et~al., \emph{{Microlensing optical
  depth towards the galactic bulge using clump giants from the MACHO survey}},
  \href{http://dx.doi.org/10.1086/432246}{\emph{Astrophys. J.} {\bf 631} (2005)
  879--905}, [\href{https://arxiv.org/abs/astro-ph/0410319}{{\tt
  astro-ph/0410319}}].

\bibitem{Bovy:2013raa}
J.~Bovy and H.-W. Rix, \emph{{A Direct Dynamical Measurement of the Milky Way's
  Disk Surface Density Profile, Disk Scale Length, and Dark Matter Profile at 4
  kpc $\stackrel{<}{\sim}$ R $\stackrel{<}{\sim}$ 9 kpc}},
  \href{http://dx.doi.org/10.1088/0004-637X/779/2/115}{\emph{Astrophys. J.}
  {\bf 779} (2013) 115}, [\href{https://arxiv.org/abs/1309.0809}{{\tt
  1309.0809}}].

\bibitem{Schoenrich:2009bx}
R.~Schoenrich, J.~Binney and W.~Dehnen, \emph{{Local Kinematics and the Local
  Standard of Rest}},
  \href{http://dx.doi.org/10.1111/j.1365-2966.2010.16253.x}{\emph{Mon. Not.
  Roy. Astron. Soc.} {\bf 403} (2010) 1829},
  [\href{https://arxiv.org/abs/0912.3693}{{\tt 0912.3693}}].

\bibitem{Reid:2014boa}
M.~J. Reid et~al., \emph{{Trigonometric Parallaxes of High Mass Star Forming
  Regions: the Structure and Kinematics of the Milky Way}},
  \href{http://dx.doi.org/10.1088/0004-637X/783/2/130}{\emph{Astrophys. J.}
  {\bf 783} (2014) 130}, [\href{https://arxiv.org/abs/1401.5377}{{\tt
  1401.5377}}].

\bibitem{blandhawthorn2016}
J.~{Bland-Hawthorn} and O.~{Gerhard}, \emph{{The Galaxy in Context: Structural,
  Kinematic, and Integrated Properties}},
  \href{http://dx.doi.org/10.1146/annurev-astro-081915-023441}{\emph{\araa}
  {\bf 54} (Sept., 2016) 529--596},
  [\href{https://arxiv.org/abs/1602.07702}{{\tt 1602.07702}}].

\bibitem{Abuter:2018drb}
{\scshape GRAVITY} collaboration, R.~Abuter et~al., \emph{{Detection of the
  gravitational redshift in the orbit of the star S2 near the Galactic centre
  massive black hole}},
  \href{http://dx.doi.org/10.1051/0004-6361/201833718}{\emph{Astron.
  Astrophys.} {\bf 615} (2018) L15},
  [\href{https://arxiv.org/abs/1807.09409}{{\tt 1807.09409}}].

\bibitem{Rolke:2004mj}
W.~A. Rolke, A.~M. Lopez and J.~Conrad, \emph{{Limits and confidence intervals
  in the presence of nuisance parameters}},
  \href{http://dx.doi.org/10.1016/j.nima.2005.05.068}{\emph{Nucl. Instrum.
  Meth.} {\bf A551} (2005) 493--503},
  [\href{https://arxiv.org/abs/physics/0403059}{{\tt physics/0403059}}].

\bibitem{Malkin:2013ac}
Z.~Malkin, \emph{{Analysis of Determinations of the Distance between the Sun
  and the Galactic Center}},
  \href{http://dx.doi.org/10.1134/S1063772913020078}{\emph{Astron. Rep.} {\bf
  57} (2013) 128--133}, [\href{https://arxiv.org/abs/1301.7011}{{\tt
  1301.7011}}].

\bibitem{Iocco:2016itg}
F.~Iocco and M.~Benito, \emph{{An estimate of the DM profile in the Galactic
  bulge region}},
  \href{http://dx.doi.org/10.1016/j.dark.2016.12.004}{\emph{Phys. Dark Univ.}
  {\bf 15} (2017) 90--95}, [\href{https://arxiv.org/abs/1611.09861}{{\tt
  1611.09861}}].

\bibitem{Karukes2019}
E.~{Karukes}, M.~{Benito}, F.~{Iocco}, R.~{Trotta} and A.~{Geringer-Sameth},
  \emph{{A Bayesian determination of the Milky Way dark matter distribution}},
  {\emph{in preparation} (2019) }.

\bibitem{Daylan:2014rsa}
T.~Daylan, D.~P. Finkbeiner, D.~Hooper, T.~Linden, S.~K.~N. Portillo, N.~L.
  Rodd et~al., \emph{{The characterization of the gamma-ray signal from the
  central Milky Way: A case for annihilating dark matter}},
  \href{http://dx.doi.org/10.1016/j.dark.2015.12.005}{\emph{Phys. Dark Univ.}
  {\bf 12} (2016) 1--23}, [\href{https://arxiv.org/abs/1402.6703}{{\tt
  1402.6703}}].

\bibitem{Abazajian:2014fta}
K.~N. Abazajian, N.~Canac, S.~Horiuchi and M.~Kaplinghat, \emph{{Astrophysical
  and Dark Matter Interpretations of Extended Gamma-Ray Emission from the
  Galactic Center}},
  \href{http://dx.doi.org/10.1103/PhysRevD.90.023526}{\emph{Phys. Rev.} {\bf
  D90} (2014) 023526}, [\href{https://arxiv.org/abs/1402.4090}{{\tt
  1402.4090}}].

\bibitem{Gordon:2013vta}
C.~Gordon and O.~Macias, \emph{{Dark Matter and Pulsar Model Constraints from
  Galactic Center Fermi-LAT Gamma Ray Observations}},
  \href{http://dx.doi.org/10.1103/PhysRevD.88.083521,
  10.1103/PhysRevD.89.049901}{\emph{Phys. Rev.} {\bf D88} (2013) 083521},
  [\href{https://arxiv.org/abs/1306.5725}{{\tt 1306.5725}}].

\bibitem{TheFermi-LAT:2015kwa}
{\scshape Fermi-LAT} collaboration, M.~Ajello et~al., \emph{{Fermi-LAT
  Observations of High-Energy $\gamma$-Ray Emission Toward the Galactic
  Center}},
  \href{http://dx.doi.org/10.3847/0004-637X/819/1/44}{\emph{Astrophys. J.} {\bf
  819} (2016) 44}, [\href{https://arxiv.org/abs/1511.02938}{{\tt 1511.02938}}].

\bibitem{Petrovic:2014xra}
J.~Petrovic, P.~D. Serpico and G.~Zaharijas, \emph{{Millisecond pulsars and the
  Galactic Center gamma-ray excess: the importance of luminosity function and
  secondary emission}},
  \href{http://dx.doi.org/10.1088/1475-7516/2015/02/023}{\emph{JCAP} {\bf 1502}
  (2015) 023}, [\href{https://arxiv.org/abs/1411.2980}{{\tt 1411.2980}}].

\bibitem{Cholis:2015dea}
I.~Cholis, C.~Evoli, F.~Calore, T.~Linden, C.~Weniger and D.~Hooper, \emph{{The
  Galactic Center GeV Excess from a Series of Leptonic Cosmic-Ray Outbursts}},
  \href{http://dx.doi.org/10.1088/1475-7516/2015/12/005}{\emph{JCAP} {\bf 1512}
  (2015) 005}, [\href{https://arxiv.org/abs/1506.05119}{{\tt 1506.05119}}].

\bibitem{Bartels:2015aea}
R.~Bartels, S.~Krishnamurthy and C.~Weniger, \emph{{Strong support for the
  millisecond pulsar origin of the Galactic center GeV excess}},
  \href{http://dx.doi.org/10.1103/PhysRevLett.116.051102}{\emph{Phys. Rev.
  Lett.} {\bf 116} (2016) 051102},
  [\href{https://arxiv.org/abs/1506.05104}{{\tt 1506.05104}}].

\bibitem{Lee:2015fea}
S.~K. Lee, M.~Lisanti, B.~R. Safdi, T.~R. Slatyer and W.~Xue, \emph{{Evidence
  for Unresolved $\gamma$-Ray Point Sources in the Inner Galaxy}},
  \href{http://dx.doi.org/10.1103/PhysRevLett.116.051103}{\emph{Phys. Rev.
  Lett.} {\bf 116} (2016) 051103},
  [\href{https://arxiv.org/abs/1506.05124}{{\tt 1506.05124}}].

\bibitem{Cirelli:2010xx}
M.~Cirelli, G.~Corcella, A.~Hektor, G.~Hutsi, M.~Kadastik, P.~Panci et~al.,
  \emph{{PPPC 4 DM ID: A Poor Particle Physicist Cookbook for Dark Matter
  Indirect Detection}}, \href{http://dx.doi.org/10.1088/1475-7516/2012/10/E01,
  10.1088/1475-7516/2011/03/051}{\emph{JCAP} {\bf 1103} (2011) 051},
  [\href{https://arxiv.org/abs/1012.4515}{{\tt 1012.4515}}].

\bibitem{Cuoco:2016jqt}
A.~Cuoco, B.~Eiteneuer, J.~Heisig and M.~Kraemer, \emph{{A global fit of the
  $\gamma$-ray galactic center excess within the scalar singlet Higgs portal
  model}}, \href{http://dx.doi.org/10.1088/1475-7516/2016/06/050}{\emph{JCAP}
  {\bf 1606} (2016) 050}, [\href{https://arxiv.org/abs/1603.08228}{{\tt
  1603.08228}}].

\bibitem{Ackermann:2015zua}
{\scshape Fermi-LAT} collaboration, M.~Ackermann et~al., \emph{{Searching for
  Dark Matter Annihilation from Milky Way Dwarf Spheroidal Galaxies with Six
  Years of Fermi Large Area Telescope Data}},
  \href{http://dx.doi.org/10.1103/PhysRevLett.115.231301}{\emph{Phys. Rev.
  Lett.} {\bf 115} (2015) 231301},
  [\href{https://arxiv.org/abs/1503.02641}{{\tt 1503.02641}}].

\bibitem{Fermi-LAT:2016uux}
{\scshape DES, Fermi-LAT} collaboration, A.~Albert et~al., \emph{{Searching for
  Dark Matter Annihilation in Recently Discovered Milky Way Satellites with
  Fermi-LAT}},
  \href{http://dx.doi.org/10.3847/1538-4357/834/2/110}{\emph{Astrophys. J.}
  {\bf 834} (2017) 110}, [\href{https://arxiv.org/abs/1611.03184}{{\tt
  1611.03184}}].

\bibitem{Bonnivard:2015xpq}
V.~Bonnivard et~al., \emph{{Dark matter annihilation and decay in dwarf
  spheroidal galaxies: The classical and ultrafaint dSphs}},
  \href{http://dx.doi.org/10.1093/mnras/stv1601}{\emph{Mon. Not. Roy. Astron.
  Soc.} {\bf 453} (2015) 849--867},
  [\href{https://arxiv.org/abs/1504.02048}{{\tt 1504.02048}}].

\bibitem{Calore:2018sdx}
F.~Calore, P.~D. Serpico and B.~Zaldivar, \emph{{Dark matter constraints from
  dwarf galaxies: a data-driven analysis}},
  \href{https://arxiv.org/abs/1803.05508}{{\tt 1803.05508}}.

\bibitem{Mazziotta:2012ux}
M.~N. Mazziotta, F.~Loparco, F.~de~Palma and N.~Giglietto, \emph{{A
  model-independent analysis of the Fermi Large Area Telescope gamma-ray data
  from the Milky Way dwarf galaxies and halo to constrain dark matter
  scenarios}},
  \href{http://dx.doi.org/10.1016/j.astropartphys.2012.07.005}{\emph{Astropart.
  Phys.} {\bf 37} (2012) 26--39}, [\href{https://arxiv.org/abs/1203.6731}{{\tt
  1203.6731}}].

\bibitem{Benito:2016kyp}
M.~Benito, N.~Bernal, N.~Bozorgnia, F.~Calore and F.~Iocco, \emph{{Particle
  Dark Matter Constraints: the Effect of Galactic Uncertainties}},
  \href{http://dx.doi.org/10.1088/1475-7516/2017/02/007,
  10.1088/1475-7516/2018/06/E01}{\emph{JCAP} {\bf 1702} (2017) 007},
  [\href{https://arxiv.org/abs/1612.02010}{{\tt 1612.02010}}].

\bibitem{Cuoco:2017rxb}
A.~Cuoco, J.~Heisig, M.~Korsmeier and M.~Kraemer, \emph{{Probing dark matter
  annihilation in the Galaxy with antiprotons and gamma rays}},
  \href{http://dx.doi.org/10.1088/1475-7516/2017/10/053}{\emph{JCAP} {\bf 1710}
  (2017) 053}, [\href{https://arxiv.org/abs/1704.08258}{{\tt 1704.08258}}].

\bibitem{Abazajian:2015raa}
K.~N. Abazajian and R.~E. Keeley, \emph{{Bright gamma-ray Galactic Center
  excess and dark dwarfs: Strong tension for dark matter annihilation despite
  Milky Way halo profile and diffuse emission uncertainties}},
  \href{http://dx.doi.org/10.1103/PhysRevD.93.083514}{\emph{Phys. Rev.} {\bf
  D93} (2016) 083514}, [\href{https://arxiv.org/abs/1510.06424}{{\tt
  1510.06424}}].

\bibitem{Keeley:2017fbz}
R.~Keeley, K.~Abazajian, A.~Kwa, N.~Rodd and B.~Safdi, \emph{{What the Milky
  WayÕs dwarfs tell us about the Galactic Center extended gamma-ray excess}},
  \href{http://dx.doi.org/10.1103/PhysRevD.97.103007}{\emph{Phys. Rev.} {\bf
  D97} (2018) 103007}, [\href{https://arxiv.org/abs/1710.03215}{{\tt
  1710.03215}}].

\bibitem{Cerdeno:2010jj}
D.~G. Cerdeno and A.~M. Green, \emph{{Direct detection of WIMPs}},
  \href{https://arxiv.org/abs/1002.1912}{{\tt 1002.1912}}.

\bibitem{2018arXiv180512562A}
E.~{Aprile}, J.~{Aalbers}, F.~{Agostini}, M.~{Alfonsi}, L.~{Althueser}, F.~D.
  {Amaro} et~al., \emph{{Dark Matter Search Results from a One
  Tonne$\times$Year Exposure of XENON1T}}, {\emph{ArXiv e-prints} (May, 2018)
  }, [\href{https://arxiv.org/abs/1805.12562}{{\tt 1805.12562}}].

\bibitem{Ibarra:2018yxq}
A.~Ibarra, B.~J. Kavanagh and A.~Rappelt, \emph{{Bracketing the impact of
  astrophysical uncertainties on local dark matter searches}},
  \href{https://arxiv.org/abs/1806.08714}{{\tt 1806.08714}}.

\bibitem{Stanek1997}
K.~Z. {Stanek}, A.~{Udalski}, M.~{Szyma{\'n}ski}, J.~{Ka{\l}u{\.z}ny}, Z.~M.
  {Kubiak}, M.~{Mateo} et~al., \emph{{Modeling the Galactic Bar Using Red Clump
  Giants}}, \href{http://dx.doi.org/10.1086/303702}{\emph{\apj} {\bf 477}
  (Mar., 1997) 163--175}, [\href{https://arxiv.org/abs/astro-ph/9605162}{{\tt
  astro-ph/9605162}}].

\bibitem{HanGould2003}
C.~{Han} and A.~{Gould}, \emph{{Stellar Contribution to the Galactic Bulge
  Microlensing Optical Depth}},
  \href{http://dx.doi.org/10.1086/375706}{\emph{\apj} {\bf 592} (July, 2003)
  172--175}, [\href{https://arxiv.org/abs/astro-ph/0303309}{{\tt
  astro-ph/0303309}}].

\bibitem{ForemanMackey:2012ig}
D.~Foreman-Mackey, D.~W. Hogg, D.~Lang and J.~Goodman, \emph{{emcee: The MCMC
  Hammer}}, \href{http://dx.doi.org/10.1086/670067}{\emph{Publ. Astron. Soc.
  Pac.} {\bf 125} (2013) 306--312},
  [\href{https://arxiv.org/abs/1202.3665}{{\tt 1202.3665}}].

\bibitem{Korsmeier:2016kha}
M.~Korsmeier and A.~Cuoco, \emph{{Galactic cosmic-ray propagation in the light
  of AMS-02: Analysis of protons, helium, and antiprotons}},
  \href{http://dx.doi.org/10.1103/PhysRevD.94.123019}{\emph{Phys. Rev.} {\bf
  D94} (2016) 123019}, [\href{https://arxiv.org/abs/1607.06093}{{\tt
  1607.06093}}].

\end{thebibliography}\endgroup

\vspace{1.5cm}
\section*{Appendix A: Additional Tests}

In this appendix we study the effect of different binning and fitting methods on the results presented in the main text.

\subsection*{($x$,$v$) vs ($x$,$\omega$) fit}

\begin{figure}[t]
\centering
\includegraphics[width=0.65\columnwidth]{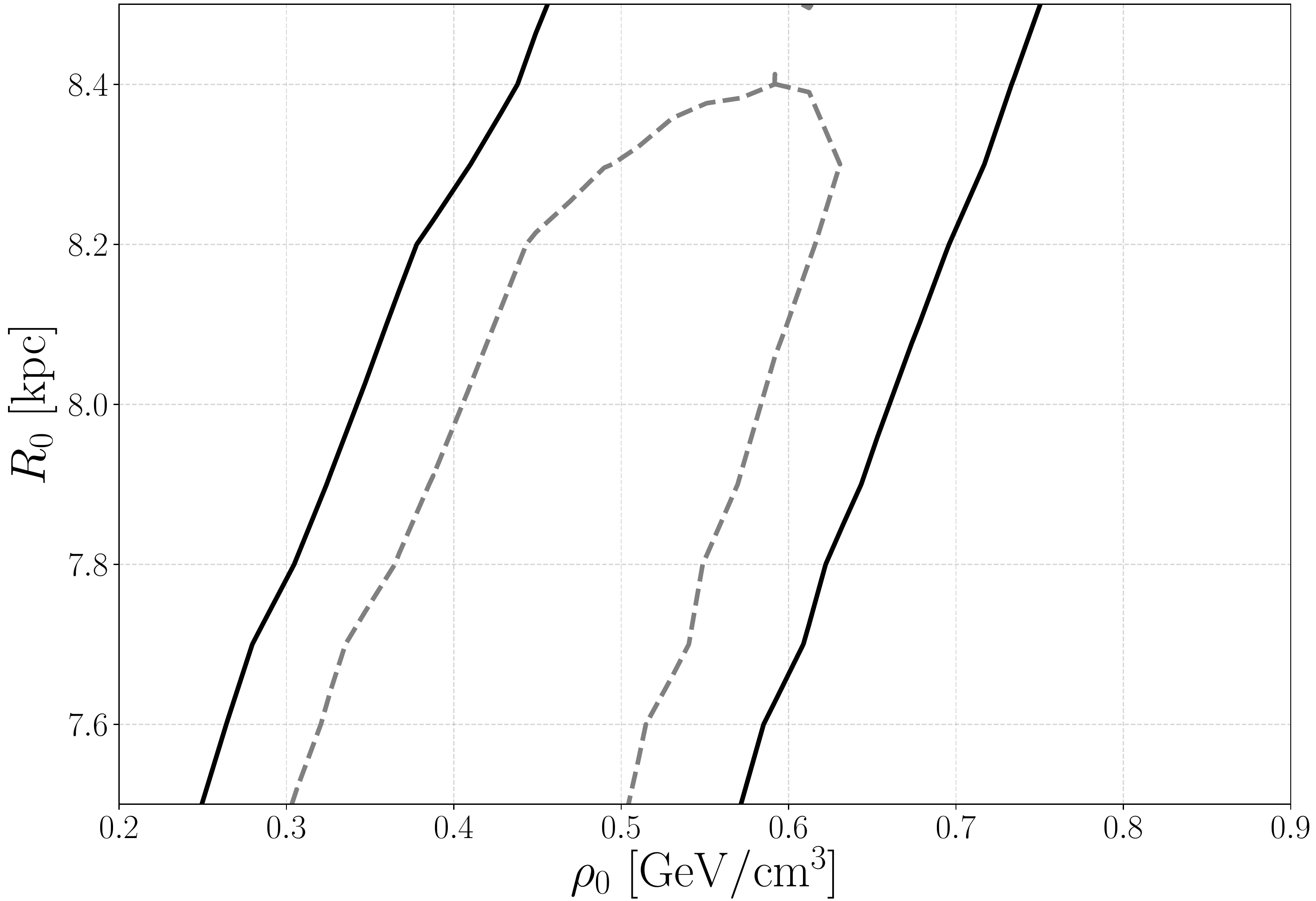}
\caption{1-2 $\sigma$ contours in the $(\rho_0, R_0)$ plane profiled over morphology and for fixed $\gamma=1$ when using ($x$,$v$) both in the binning and fitting.}
\label{fig:contours_xv}
\end{figure}

As explained in section~\ref{sec:binning}, we perform the main fit in the ($x$,$\omega$) plane
where the unbinned data-points have uncorrelated $x$ and $\omega $
uncertainties. Here, we test the effect of fitting, instead, in the ($x$,$v$) plane.
Unbinned data in the ($x$,$v$) plane   are binned according to the procedure described in section~\ref{sec:binning}.
Results are shown in Fig.\ref{fig:contours_xv} for the example case of the $(\rho_0, R_0)$ plane .
It can be seen that the results of the ($x$,$v$) fit are compatible with the ($x$,$\omega$) fit,
although some differences can be seen, as a slight shift toward lower $\rho_0$ values
of about 0.05 GeV/cm$^3$, which is anyway small with respect to the overall width of the contours,
and slightly different slope of the $(\rho_0, R_0)$ degeneracy.
Despite the overall agreement, the small differences among the two fits suggest
nonetheless that performing the analysis in the ($x$,$\omega$)  plane is
a more robust procedure since the properties of the errors (i.e., uncorrelated) are more straightforward.

\subsection*{Fit with a larger number of bins}

\begin{figure}[t]
\centering
\includegraphics[width=0.65\columnwidth]{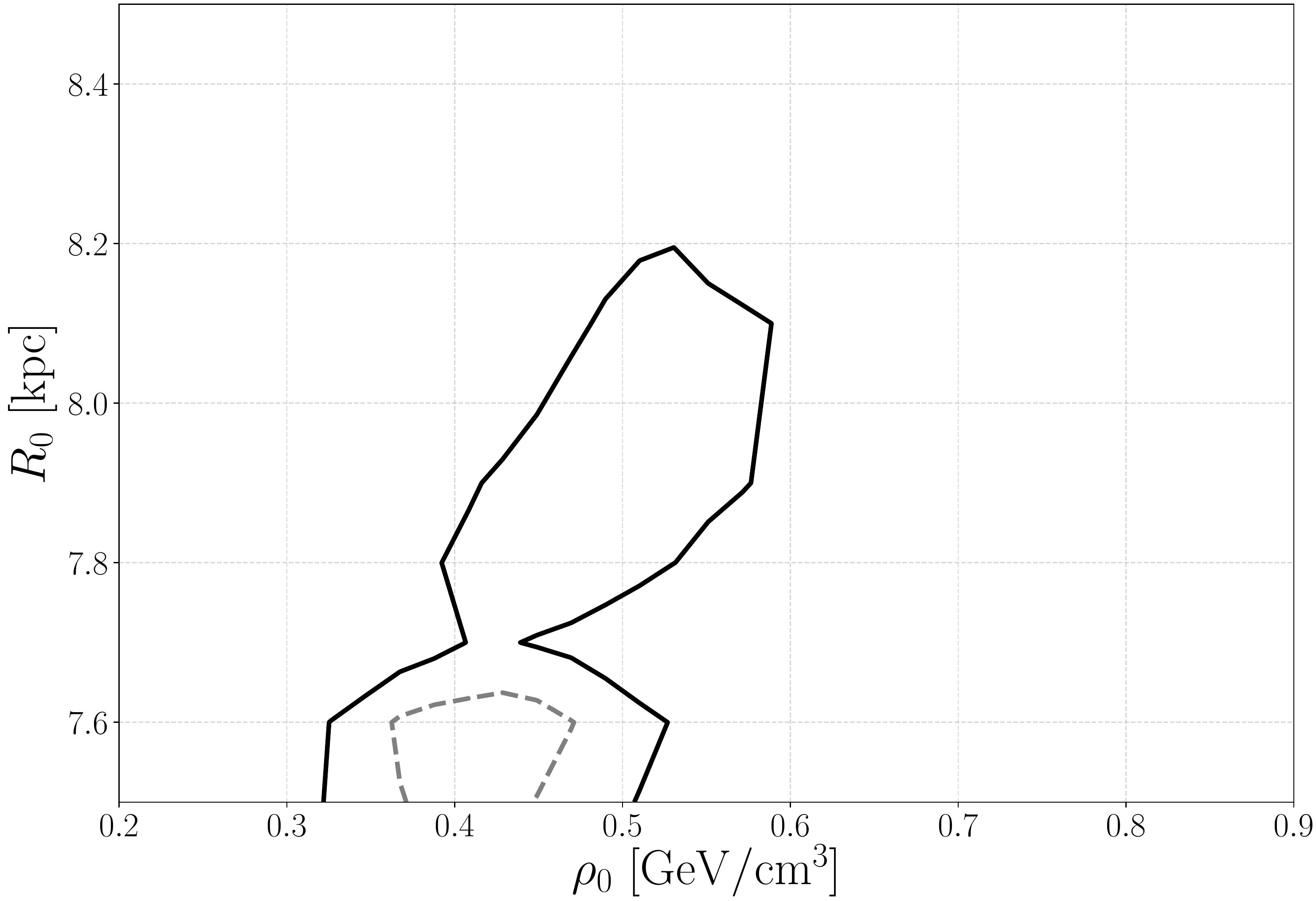}
\caption{1-2 $\sigma$ contours in the $(\rho_0, R_0)$ plane profiled over morphology and for fixed $\gamma=1$ when using ($x$,$\omega$) with 54 bins both in the binning and fitting.}
\label{fig:contours_xy54}
\end{figure}

We have, furthermore, tested the effect of increasing the number of bins used in the fit in the default $(x, \omega)$ procedure.
In the present case the bins are chosen again starting from $x=2.5/8$  and ending at $x=22/8$, but we take
i) 40 bins (up to 10.5/8) with a width of $\Delta x=0.2/8$ (up to 10.5/8),  
ii) 9 bins with $\Delta x$=0.5/8 (up to 15/8),
iii) 3 bins with $\Delta x=1/8.0$ (up to 18/8) and
iv) 2 last bins with  $\Delta x = 2/8$,
for  a total of 54 bins. This is roughly double with respect to the default setup, which has 25 bins.
In each bin we use the procedure outlined in section~\ref{sec:binning} to derive the 
central value and the error.
Results of the fit for this case are shown in Fig.~\ref{fig:contours_xy54} for the example case of the $(\rho_0, R_0)$ plane.
As expected, the main effect is a reduction of the errors, as also discussed in section \ref{sec:comparison} and in ref.~\cite{Pato:2015dua}.
In particular,  the allowed range for $\rho_0$ at 2$\sigma$ is now in the range \mbox{$0.3.- 0.6$ GeV/cm$^3$}.
The analysis also becomes more sensitive to $R_0$, and values above 8.2 kpc    are disfavored at  2$\sigma$.

\subsection*{Frequentist vs Bayesian}

\begin{figure}[t]
\centering
\includegraphics[width=0.65\columnwidth]{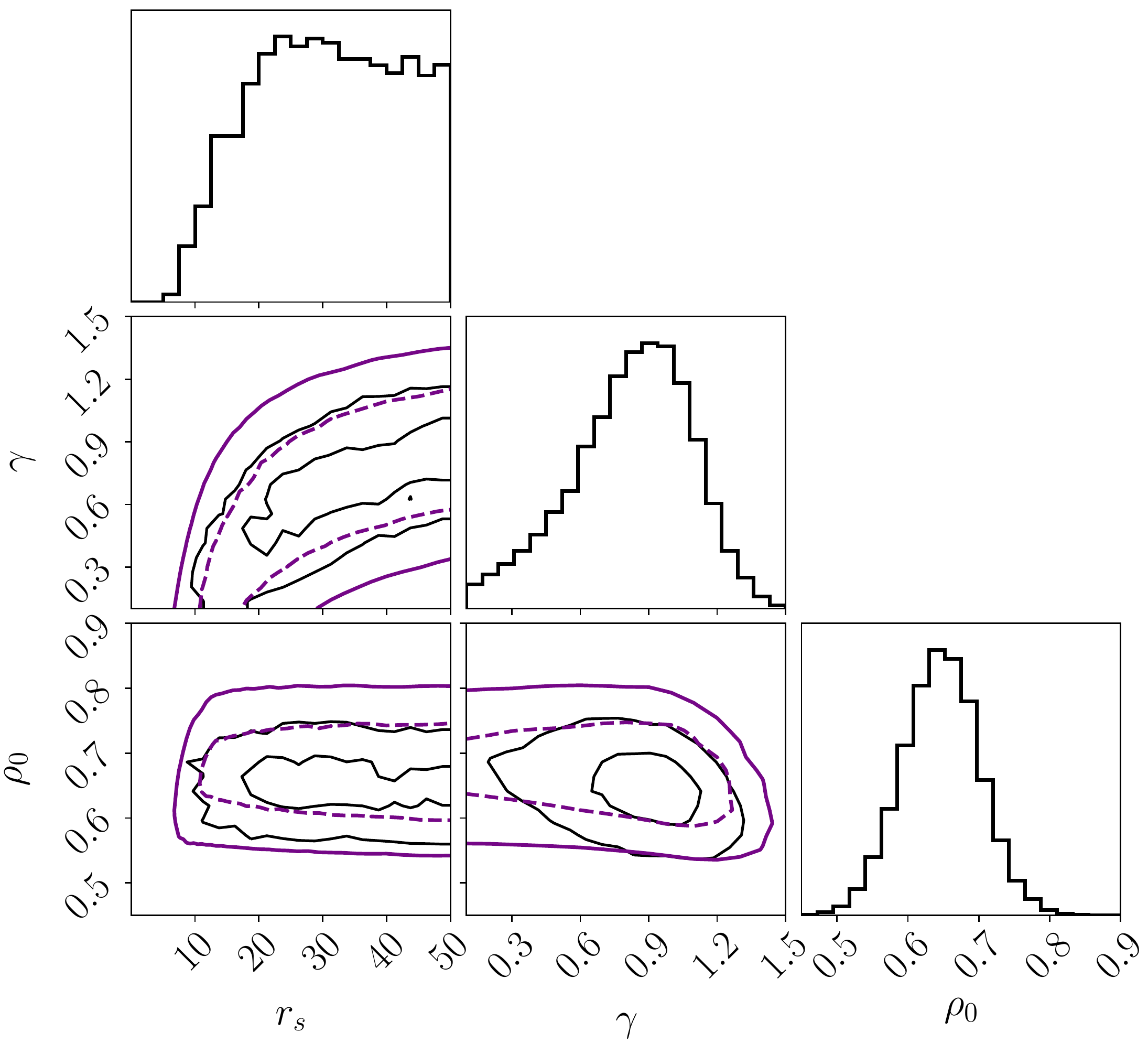}
\caption{Triangle plot comparing the results of a Bayesian (black lines) vs frequentist (purple lines) fit of a five-dimensional analysis
in  $\gamma$, $R_s$ and $\rho_0$,  $\langle\tau\rangle$, $\Sigma_*$. The last two parameters are not shown. 
Contours are at 1 and 2 $\sigma$. See text for more details.}
\label{fig:BayesvsFreq}
\end{figure}

Another possible uncertainty is given by the use of the statistical methodology.
To test this effect we compare our default methodology, which makes use of a grid
in the parameter space and frequentist formalism, which a fully Bayesian analysis.
To this purpose we use a simplified framework where we fix $R_0=8.34$ kpc and
a single baryonic morphology, in particular the one labeled bJ (which assumes the {\it E2} bulge given in \cite{Stanek1997} and the stellar disc from \cite{HanGould2003}).
In this case, we thus have only five parameters, instead of the seven ones explored in the main analysis.
To perform the Bayesian analysis we use a Monte Carlo scan of the parameter space
with the \emph{emcee} tool \cite{ForemanMackey:2012ig}, and use flat priors on the parameters.
The results are shown in the triangle plot of Fig.~\ref{fig:BayesvsFreq}.
The purple lines show the 1 and 2 $\sigma$ frequentist contours build with the method
described in the main text, while the black lines show the analogous Bayesian result.
The triangle diagonal shows the 1d Bayesian posterior for the single parameters.
The triangle plot focus on the three DM Halo parameters $\gamma$, $R_s$ and $\rho_0$,
and it does not show $\langle\tau\rangle$, $\Sigma_*$, although the fit is five-dimensional.
As can be seen the Bayesian and frequentist contours are in excellent agreement. 
The only clear difference is that the frequentist contours are slightly larger,
and thus more conservative. This is a typical result, especially when some of the parameters
is not well constrained, as in this case. In the case when all the parameters are well
constrained typically the agreement between the two methods is even closer (for example, see \cite{Korsmeier:2016kha}.)

\section*{Appendix B: Burkert Profile}

\begin{figure}[t]
\centering
\includegraphics[width=0.48\columnwidth]{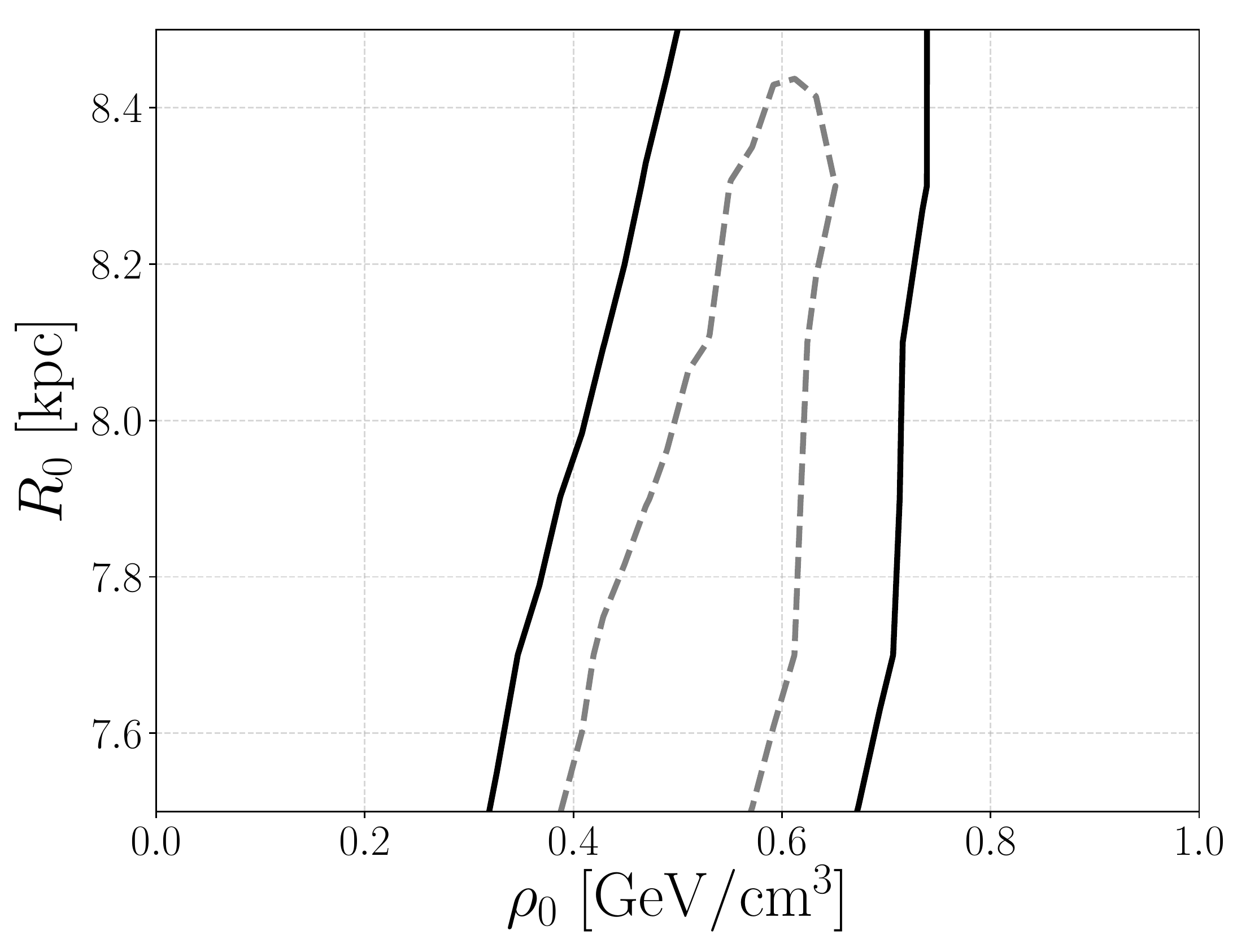}
\includegraphics[width=0.48\columnwidth]{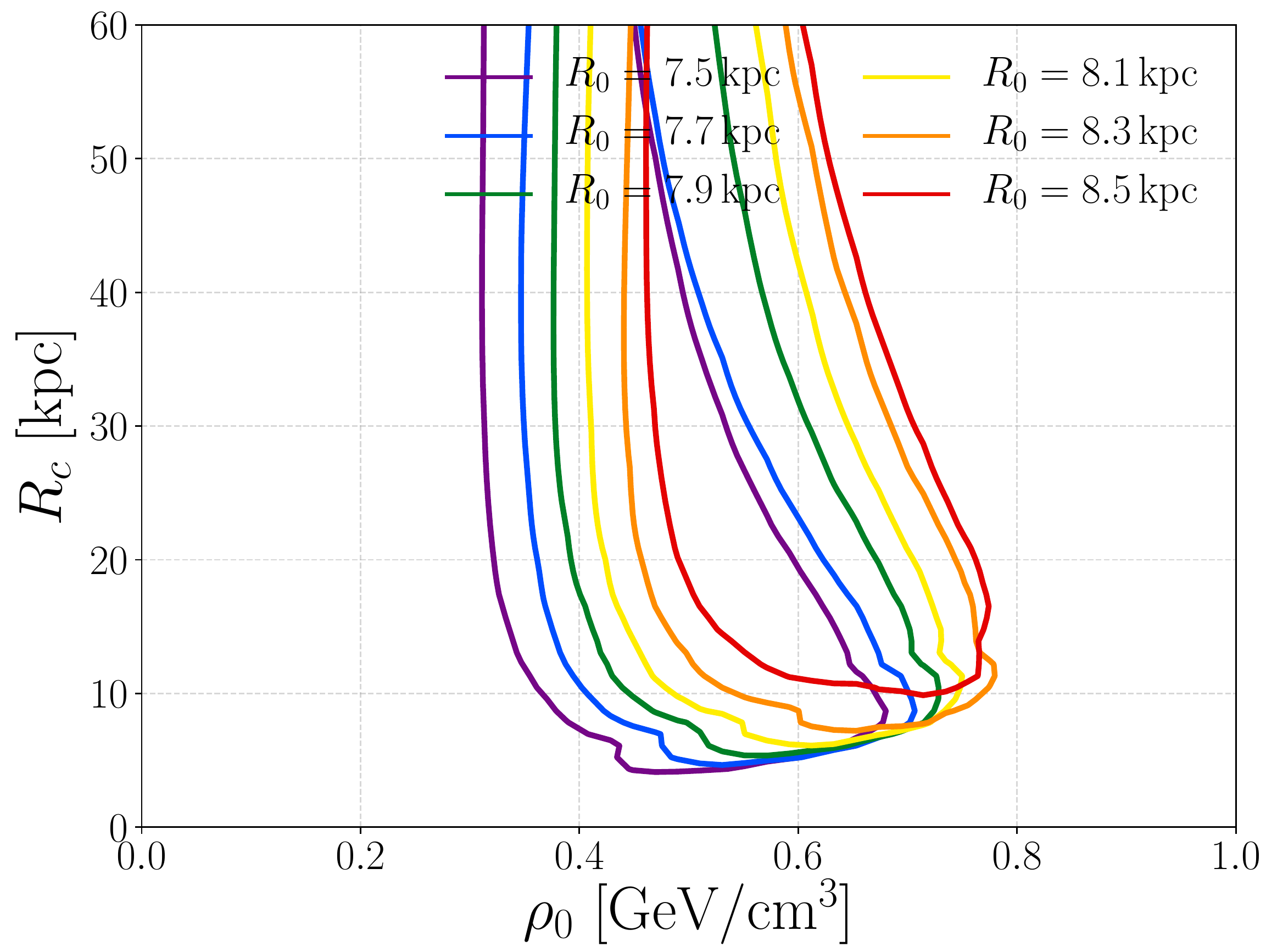}
\caption{Left panel: 1-2$\sigma$ contours in the $(\rho_0,R_0)$ plane profiled over $R_c$, $\mathcal{M}_i$, $\langle\tau\rangle$ and $\Sigma_*$.
Right Panel:  2$\sigma$ contours in the $(R_c,\rho_0)$ plane
 for  various values of $R_0$ and profiled over $\mathcal{M}_i$, $\langle\tau\rangle$ and $\Sigma_*$.}
\label{fig:BUR}
\end{figure}

In this appendix we discuss the results of the fit when the Burkert (BUR) profile~\cite{Burkert:1995yz} is adopted, i.e.,
\begin{equation}
  \rho(r)=  \rho_s     \left( \frac{r}{R_c}\right)^{-1}   \left(  1 + \frac{r^2}{R_c^2} \right)^{-1} \ ,
\end{equation}
 instead of the gNFW used in the main text.
A simplification in this case arises from the fact that only two parameters define the model, i.e., $R_c$ the core radius, and, $\rho_s$
the scale density, instead of the three of the gNFW case.
The left panel of Fig.~\ref{fig:BUR} is the BUR analogue of the lower-right panel of Fig.~\ref{fig:contours_super_allR0} in the main text.
The results for BUR and gNFW are fully compatible, with the BUR case giving a slightly tighter degeneracy between $R_0$ and $\rho_0$.
The right panel shows the $(R_c,\rho_0)$ plane. An emerging interesting feature is that a minimum core size of about $\sim 5$ kpc is present. 
This appears to be a peculiarity of the BUR profile, while smaller core sizes should be possible if different profile
parameterization are employed.
The best-fit $\chi^2$ for the BUR case is $\sim 6$, similar to the gNFW case, indicating the two profiles can
provide equally good fits to the Galactic rotation curve.
As for the gNFW case, the tabulated likelihood in $R_0, R_c, \rho_s$, profiled over  $\mathcal{M}_i$, $\langle\tau\rangle$ and $\Sigma_*$
is provided at \href{https://github.com/mariabenitocst/UncertaintiesDMinTheMW}{https://github.com/mariabenitocst/UncertaintiesDMinTheMW}.

\begin{figure}[t]
\centering
\includegraphics[width=0.45\columnwidth]{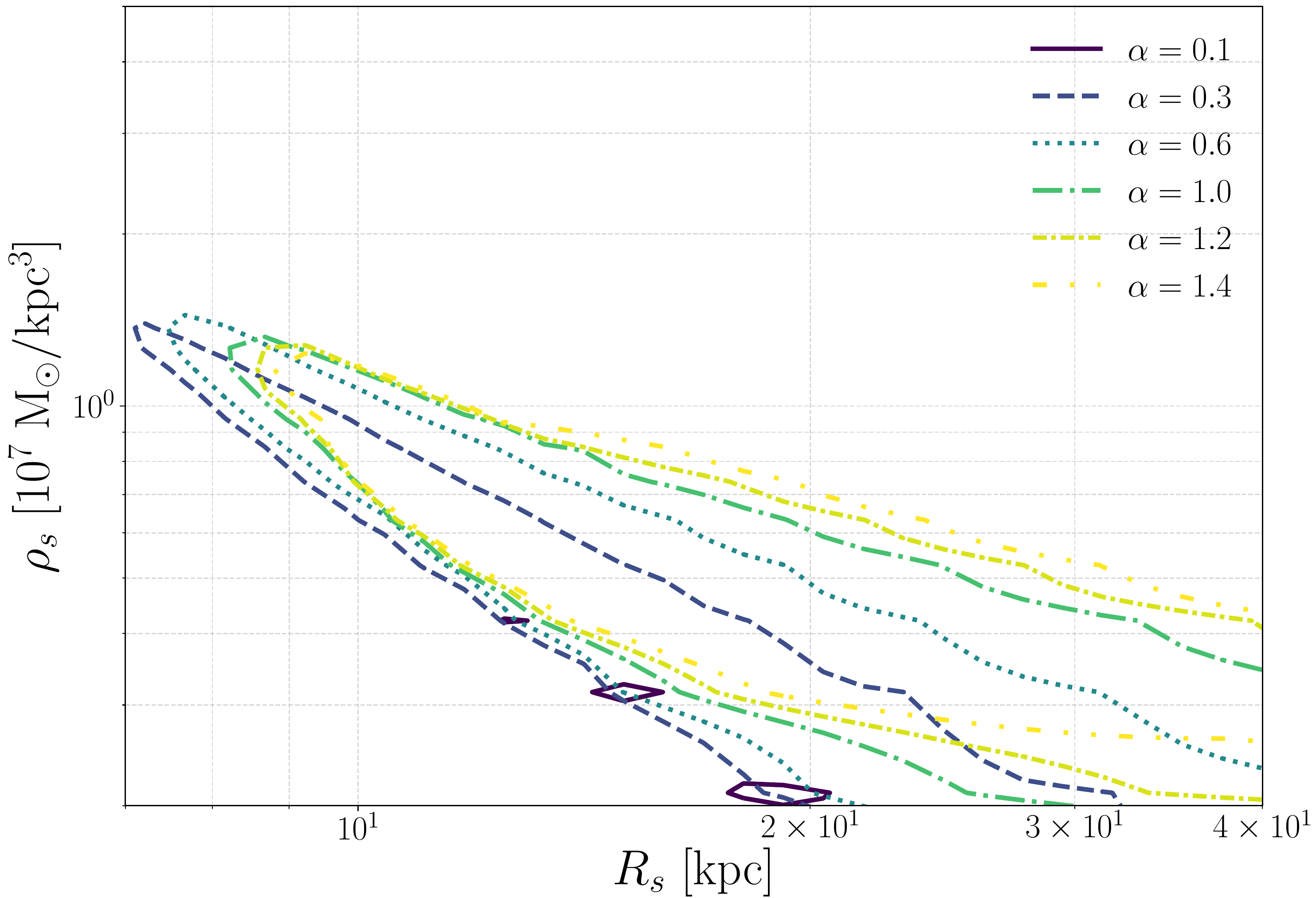}
\includegraphics[width=0.45\columnwidth]{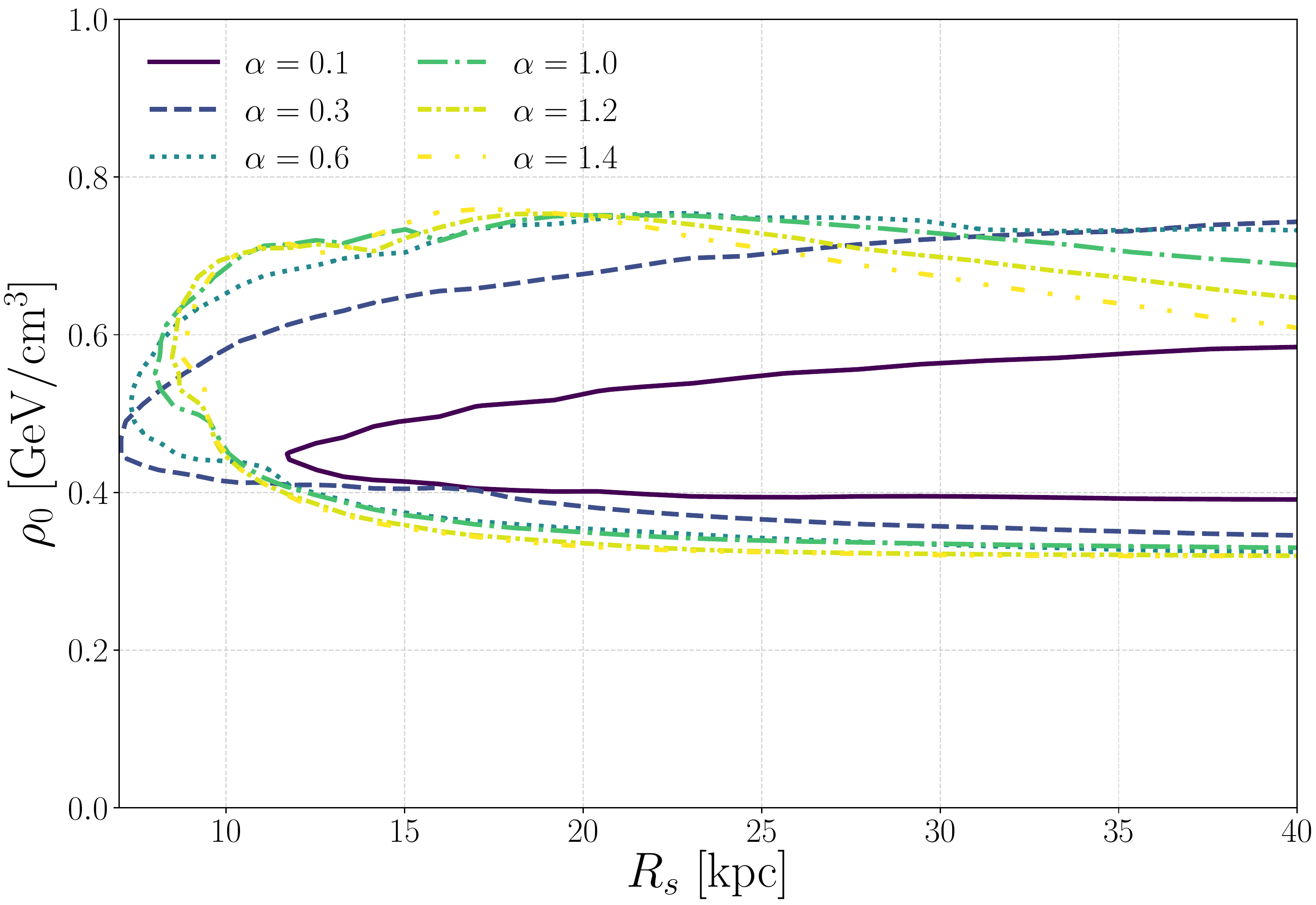}
\includegraphics[width=0.45\columnwidth]{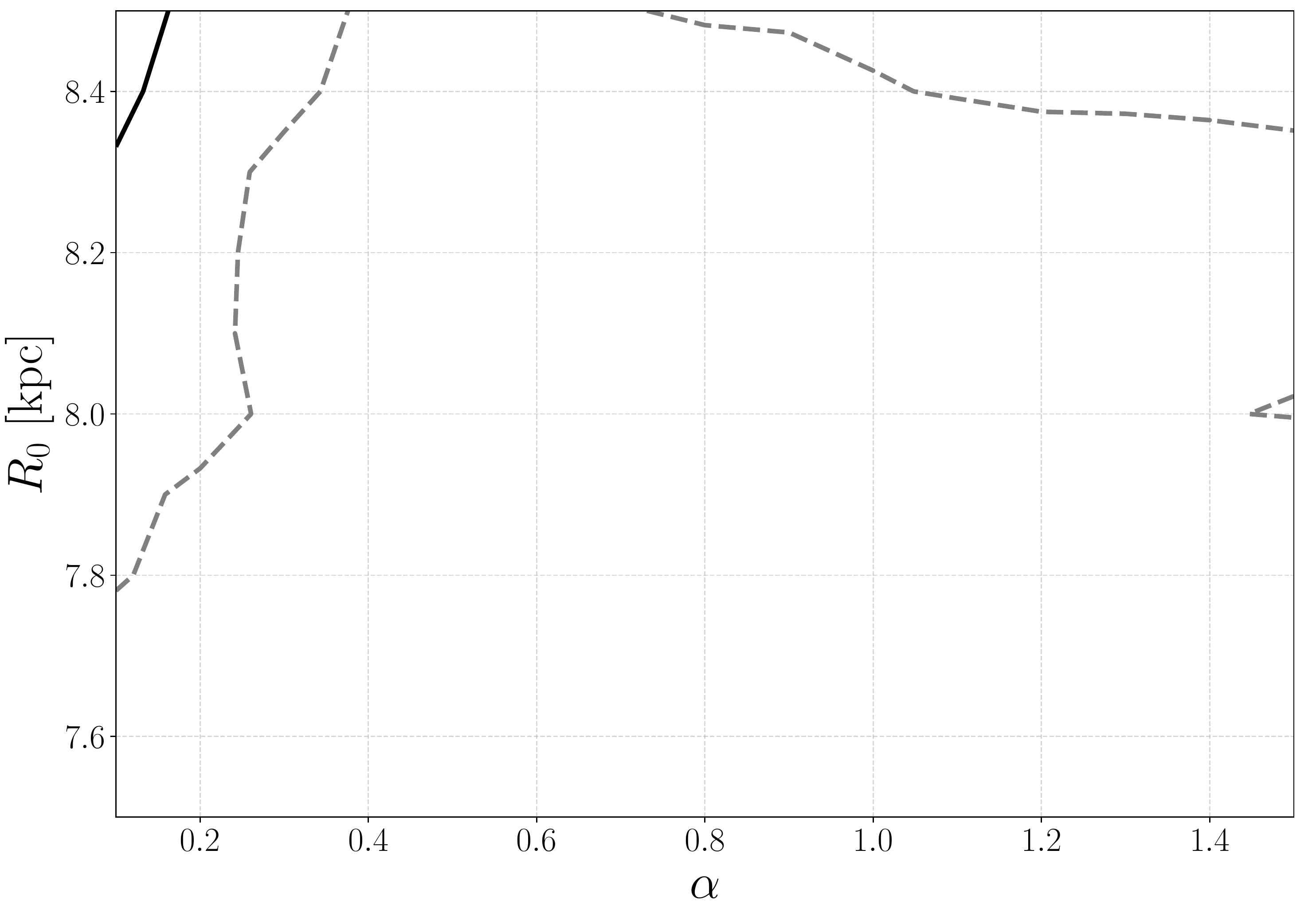}
\includegraphics[width=0.45\columnwidth]{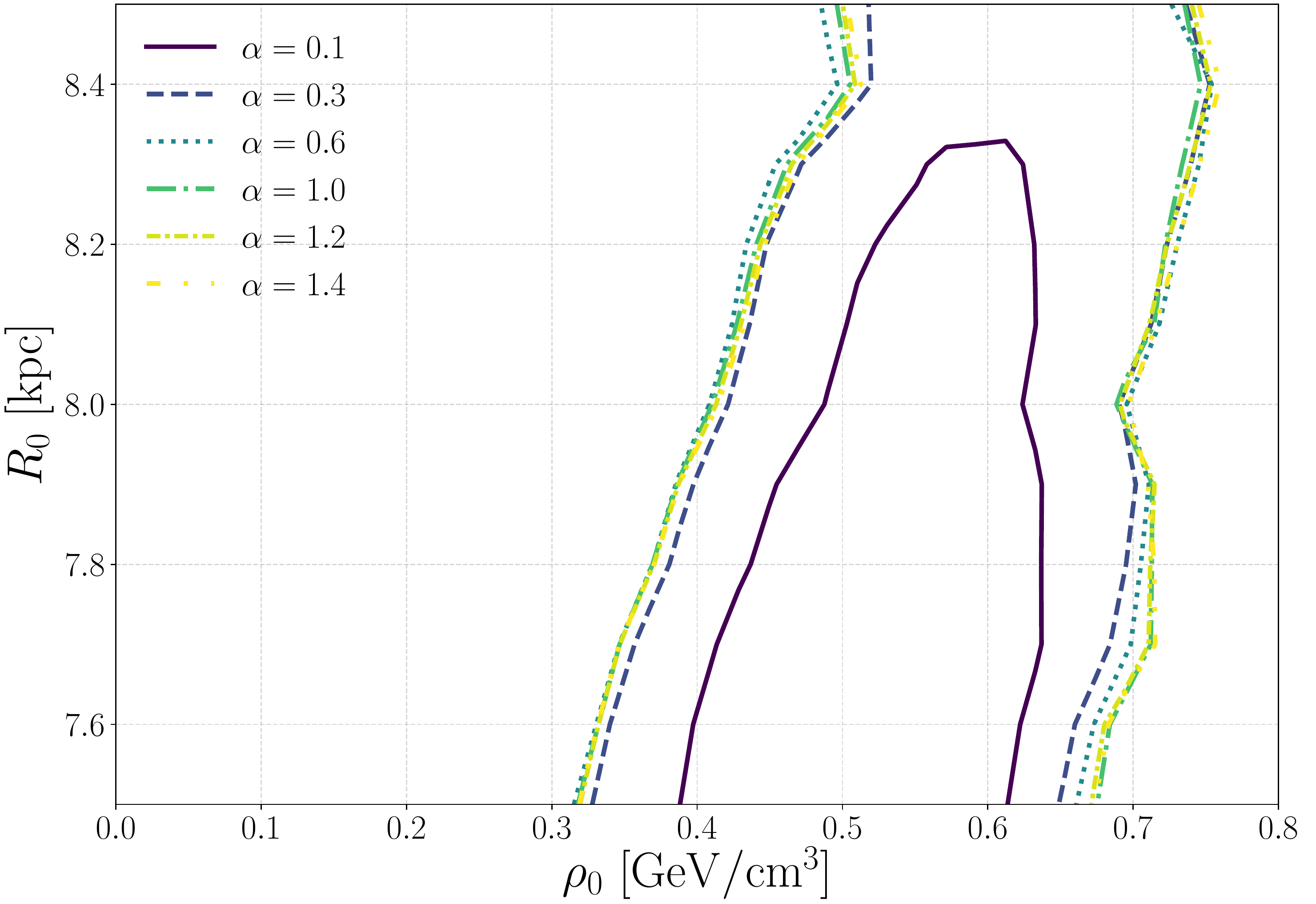}
\caption{Upper-left panel: 2$\sigma$ contours in the $(R_s,  \rho_s)$ plane for
 various fixed values of $\alpha$ profiled over $\mathcal{M}_i$, $\langle\tau\rangle$, $\Sigma_*$ and $R_0$.
Upper-right: same as upper-left but in the $(R_s, \rho_0)$ plane. 
Lower-left: 1-2 $\sigma$ contours of the $\chi^2$  in the $\alpha, R_0$ plane profiled over $\mathcal{M}_i$, $\langle\tau\rangle$, $\Sigma_*$, $R_s$, and  $\rho_s$.
Lower-right: 2-$\sigma$ contours of the profile $\chi^2$ in the $(\rho_0, R_0)$ plane for different fixed values of $\alpha$.}
\label{fig:contour_Rsrhos_Rsrho0_alpha_einasto}
\end{figure}

\section*{Appendix C: Einasto Profile}

Finally, we also derive results for another commonly employed profiled, i.e., the Einasto profile ~\cite{Einasto1965} ,
\begin{equation}
  \rho(r)=  \rho_s    \exp  \left( -\frac{2}{\alpha}   \left[  \left(  \frac{r}{R_s}  \right)^{\alpha} -1 \right] \right)  \ ,
\end{equation}
which is defined in terms of $R_s$, $\rho_s$ and $\alpha$, which is a shape parameters which plays
a role similar to $\gamma$ for gNFW case, although with `opposite' values, i.e., when $\alpha=1$ the profile
is cored, while small values of $\alpha$ give a cuspy profile.
Fig.~\ref{fig:contour_Rsrhos_Rsrho0_alpha_einasto} is the analogue of Fig.~\ref{fig:contour_Rsrhos_Rsrho0_gamma}
in the main text, and it shows that the same degeneracies of the gNFW profile are present for the Einasto one.
In particular, both cuspy ($\alpha \ll 1$) and cored ($\alpha=1$) profiles are compatible with the data.
The best-fit $\chi^2$ for the Einasto case is $\sim 6$, similar to the gNFW and BUR case, indicating the all the profiles considered can
provide equally good fits to the Galactic rotation curve.
As for the gNFW and BUR case, the tabulated likelihood in $R_0, R_s, \rho_s$, $\alpha$ profiled over  $\mathcal{M}_i$, $\langle\tau\rangle$ and $\Sigma_*$
is provided at \href{https://github.com/mariabenitocst/UncertaintiesDMinTheMW}{https://github.com/mariabenitocst/UncertaintiesDMinTheMW}.

\end{document}